\documentstyle[aps,prl,twocolumn,floats]{revtex}
\begin{document}
\draft
\title{{\rm PHYSICAL REVIEW E}\hfill {\sl Version of \today}\\~~\\
Random Noise and Pole-Dynamics in Unstable Front Propagation}
\author {Zeev Olami, Barak Galanti, Oleg Kupervasser and
Itamar  Procaccia}
\address{Department of~~Chemical Physics,\\
 The Weizmann Institute of Science,
Rehovot 76100, Israel}
\maketitle
\begin{abstract}
The problem of flame propagation is studied as an example of unstable fronts that
wrinkle on many scales. The analytic tool of pole expansion 
in the complex plane is employed to address the interaction of the unstable growth
process with random initial conditions and 
perturbations. We argue that the effect of random noise is immense and that it
can never be neglected in sufficiently large systems. We present simulations
that lead to scaling laws
for the velocity and acceleration of the front as a function of the system
size and the level of noise, and analytic arguments that explain these results
in terms of the noisy pole dynamics.
\end{abstract}
%\leftskip 54.8pt
%\rightskip 54.8pt
\pacs{PACS numbers 47.27.Gs, 47.27.Jv, 05.40.+j}
%}]
%\widetext
%%%%%%%%%%%%%%%%%%%%%%%%%%%%%%
\section{Introduction}
The aim of this paper is to examine the role of random fluctuations on the
dynamics
of growing wrinkled interfaces which are governed by non-linear equations
of motion.
We are interested in those examples for which the growth of a flat or smooth
interface are inherently unstable. A famous example of such growth phenomena is
provided by Laplacian growth patterns \cite{Pel,BS,Vic}. The experimental
realization of such
patterns is seen for example in Hele-Shaw cells \cite {Pel} in which air or
another low viscosity
fluid is displacing oil or some other high viscosity fluid.
Under normal conditions the advancing fronts do not remain flat; in channel
geometries
they form in time a stable finger whose width is determined by delicate effects
that arise from the existence of surface tension. In radial geometry, the 
growth the interface
forms a contorted and ramified fractal shape. A related phenomenon has been
studied
in a model equation for flame propagation which has the same linear stability
properties as the Laplacian growth problem \cite{77Siv}. The physical
problem in this case
is that of pre-mixed flames which
exist as self-sustaining fronts of exothermic chemical reactions in gaseous
combustion. Experiments \cite{89GIS} on flame propagation in radial
geometry show that the flame
front accelerates as time
goes on, and roughens with characteristic exponents. Both observations did not
receive proper theoretical explanations. It is notable that  the channel and
radial
growth are markedly different; the former leads to a single giant
cusp in the
moving front, whereas the latter exhibits infinitely many cusps that appear in
a complex hierarchy as the flame front develops \cite{94FSF,96KOP}.

Analytic techniques to study such processes are available \cite{82LC}. In
the context of
flame propagation \cite{96KOP,85TFH,89Jou,90Jou}, and in Laplacian growth in
the zero surface-tension limit \cite{84SB,86How,94DMW} one
can examine solutions that are described in terms of poles in the complex plane.
This description is very useful in providing a set of ordinary differential equations
for the positions of the poles, from which one can deduce the geometry of the
developing front in an extremely economical and efficient way.
Unfortunately this
description is not available in the case of Laplacian growth with surface
tension, and this makes the flame propagation problem very attractive.
However, it  suffers from one fundamental drawback. For the noiseless
equation the pole-dynamics always conserves the number of poles that existed in the initial
conditions.  As a result there is a final degree of ramification that is afforded
by every set of initial conditions even in the radial geometry, and it is not
obvious
how to describe the continuing self-similar growth that is seen in
experimental conditions or numerical simulations. Furthermore, as mentioned before,
at least in the
case of flame propagation one observes \cite{89GIS} an {\em acceleration}
of the flame
front with time. Such a phenomenon is impossible when the number
of poles is conserved. It is therefore tempting to conjecture that noise
may have
an important role in affecting the actual growth phenomena that are observed in
such systems. In fact, the effect of noise on unstable front dynamics has not
been adequately addressed in the literature. From the point of view of analytic
techniques noise can certainly generate new poles even if the initial conditions
had a finite number of poles. The subject of pole dynamics with the existence
of random noise, and the interaction between
random fluctuations
and deterministic front propagation are the main issues of this paper.

We opt to study the example of flame propagation rather than Laplacian growth, simply because
the former has an analytic description in terms of poles also in the experimentally
relevant
case of finite viscosity. We choose to limit the present study to channel
geometry. The reason is that in radial geometry it is more difficult to
disentangle
the effects of external noise from those of initial conditions. After all,
initially the system can contain 
 infinitely many poles,very far away near infinity in the complex
plane (and therefore having an infinitely small contribution to the interface).
Since the growth of the radius changes the stability of the system,
more and more of these poles might  fall down to the real axis and become observable. 
In channel geometry the analysis of the effect of
initial conditions
is relatively straightforward, and one can understand it before focusing on
the (more interesting) effects of external noise \cite{85TFH}.  The basic reason
for this is that
in this geometry the noiseless steady state solution for the developed front
is known analytically. As described in Section II, in a channel of width $L$
the steady-state solution is given in terms of
$N(L)$ poles that are organized on a line parallel to the imaginary axis.
It  can
be shown that for any number of poles in the initial conditions this is the only
attractor of the pole dynamics. After the establishment of this
steady state we can begin to systematically examine the effects of external
noise on this solution. As stated before, in
radial conditions there is no stable steady state with a finite number of poles,
and the disentanglement of initial vs. external perturbations is less
straightforward \cite{96KOP}.
We believe nevertheless that the insights provided in this paper have relevance
for radial growth as well as will be discussed in the final Section and in
forthcoming
publications.

We have a number of goals in this paper. Firstly, after introducing the pole
decomposition, the pole dynamics, and the basic steady state, we will
present a stability
analysis of the solutions of the flame propagation problem in a channel
geometry. It will
be shown that the giant cusp solution is linearly stable, but non-linearly
unstable. These results, which are described in Section III, can be obtained
either by
linearizing the dynamics around the giant cusp solutions in order to study the stability
eigenvalues, or by examining perturbations in the form of poles in the
complex plane. The
main result of Section III is that there exists one Goldstone mode and two
modes whose
eigenvalues hit the real axis periodically when the system size $L$
increases. Thus
the system is marginally stable at particular values of $L$, and it is always
nonlinearly unstable, allowing finite size perturbations to introduce new
poles into the system. This insight allows us to understand the relation
between the
system size and the effects of noise. In Section IV we discuss the
relaxation dynamics
that ensues after starting the system with ``small" initial data.  We
study the
coarsening process that leads in time to the final solution of the giant cusp, and
understand from this what are the typical time scales that exist in our
dynamics.
We offer in this Section some results of numerical simulations that are
interpreted
in the later sections.
In Section V we focus on the phenomenon of acceleration of the flame front
and its relation
to the existence of noise. In noiseless conditions the velocity of the flame
front in a finite channel is bounded \cite{85TFH}. This can be shown either by using the pole
dynamics or directly from the equation of motion. We will present the results
of numerical simulations where the noise is controlled, and show how the
velocity of the flame front is affected by the level of the noise and the
system size.
The main results are: (i) Noise is responsible for introducing new poles to the
system; (ii) For low levels of noise the velocity of the flame front scales
with the
system size with a characteristic exponent; (iii) There is a phase transition at
a sharp (but system-size dependent) value of the noise-level, after which the behavior
of the system changes qualitatively; (iv) After the phase transition the
velocity
of the flame front changes very rapidly with the noise level. In the last
Section we remark on the implications of these observations for the scaling behavior
of the radial growth problem, and  present a summary and conclusions.
%%%%%%%%%%%%%%%%%%%%%%%%%%%%%%%%%
\section{Equations of Motion and Pole-decomposition in the Channel Geometry}
It is known that planar flames freely propagating
through initially motionless homogeneous combustible mixtures are
intrinsically unstable.
It was reported that such flames develop characteristic structures which
include cusps,
and that under usual experimental conditions the flame front accelerates as
time goes
on. A model in $1+1$ dimensions that pertains to the propagation of flame
fronts in channels of width
$\tilde L$ was proposed in \cite{77Siv}. It is written in terms of position
$h(x,t)$ of the
flame front above the $x$-axis. After appropriate rescalings it takes
the form:
\begin{equation}
{\partial h(x,t) \over \partial t}=
{1\over 2}\left[{\partial h(x,t) \over \partial x }\right]^2
 +\nu{\partial^2 h(x,t)\over \partial x^2}+ I\{h(x,t)\}+1 \ . \label{Eqnondim}
\end{equation}
The domain is  $0 <x< \tilde L$, $\nu$ is a parameter and we use periodic boundary
conditions. The functional $I[h(x,t)]$
is the Hilbert transform which is conveniently defined in terms of the spatial
Fourier transform
\begin{eqnarray}
&&h(x,t)= \int_{-\infty}^{\infty} e^{i k x}\hat h(k,t) dk \label{Four}\\
&& I[h(k,t)] = |k| \hat h(k,t) \label{hil}
\end{eqnarray}
For the purpose of introducing the pole-decomposition it is convenient to
rescale the domain to $0<\theta <2\pi$. Performing this rescaling and
denoting the
resulting quantities with the same notation we have
\begin{eqnarray}
&&{\partial h(\theta,t) \over \partial t}=
{1\over 2L^2}\left[{\partial h(\theta,t) \over \partial \theta }\right]^2
 +{\nu\over L^2}{\partial^2 h(\theta,t)\over \partial\theta^2}\nonumber \\&&+
{1\over L}I\{h(\theta,t)\}+1 \ .
\label{Eqdim}
\end{eqnarray}
In this equation $L=\tilde L/2\pi$.
Next we change variables to $u(\theta,t)\equiv {\partial
h(\theta,t)/\partial\theta}$.
 We find
\begin{equation}
{\partial u(\theta,t) \over \partial t}=
{u(\theta,t)\over L^2}{\partial u(\theta,t) \over \partial \theta }
 +{\nu\over L^2}{\partial^2 u(\theta,t)\over \partial \theta^2}+ {1\over
L}I\{u(\theta,t)\} \ . \label{eqfinal}
\end{equation}
It is well known that the flat front solution of this equation is linearly
unstable.
The linear spectrum in $k$-representation is
\begin{equation}
\omega_k=|k|/L-\nu k^2/L^2 \ . \label{spec}
\end{equation}
There exists a
typical scale $k_{max}$ which is the last unstable mode
\begin{equation}
k_{max} = {L\over \nu} \ . \label{kmax}
\end{equation}
Nonlinear effects stabilize a new steady-state which is discussed next.

The outstanding feature of the solutions of this equation is the appearance
of cusp-like structures in the developing fronts. Therefore a representation in
terms of Fourier modes is very inefficient. Rather, it appears very 
worthwhile to represent such solutions in terms of sums of functions of poles 
in the complex plane. It will be shown below that the position of the cusp along the
front is determined by the real coordinate of the pole, whereas the height of the
cusp is in correspondence with the imaginary coordinate. Moreover, it will
be seen that the dynamics of the developing front can be usefully described in
terms of the dynamics of the poles. Following \cite{82LC,85TFH,90Jou,96KOP} we 
expand the solutions $u(\theta,t)$ in functions that depend on
$N$ poles whose position $z_j(t)\equiv x_j(t)+iy_j(t)$ in the complex plane
is time dependent:
\begin{eqnarray}
&&u(\theta,t)=\nu\sum_{j=1}^{N}\cot \left[{\theta-z_j(t) \over 2}\right]
   + c.c.\nonumber \\
&&=\nu\sum_{j=1}^{N}{2\sin [\theta-x_j(t)]\over
\cosh [y_j(t)]-\cos [\theta-x_j(t)]}\ , \label{upoles}
\end{eqnarray}
\begin{equation}
h(\theta,t)=2\nu\sum_{j=1}^{N}{\ln \Big[\cosh (y_j(t))-\cos (\theta-x_j(t))
\Big]}+C(t) \ . \label{rpoles}
\end{equation}
In (\ref{rpoles}) $C(t)$ is a function of time. The function (\ref{rpoles})
is a superposition of quasi-cusps (i.e. cusps that are rounded at the tip). The
real part of the pole position (i.e. $x_j$) is the coordinate (in the domain
$[0,2\pi]$) of the maximum
of the quasi-cusp, and the imaginary part of the pole position (i.e $y_j$)
is related to
the depth of the quasi-cusp. As $y_j$ decreases the depth of the cusp
increases. As $y_j \to 0$ the depth diverges to infinity. Conversely, when $y_j\to
\infty$ the
depth decreases to zero.

The main advantage of this representation is that the propagation and
wrinkling of the
front can be described via the dynamics of the poles. Substituting
(\ref{upoles}) in (\ref{eqfinal}) we derive the following ordinary
differential equations for the positions of the poles:
\begin{equation}
- L^2{dz_{j}\over dt}=\Big[\nu\sum_{k=1
  ,k\neq j}^{2N }\cot \left({z_j-z_k\over 2}\right)
  +i{L\over 2 }sign [Im(z_j)]\Big].\label{eqz}
\end{equation}
We note that in (\ref{upoles}), due to the complex conjugation, we have $2N$ poles 
which are arranged in pairs such that for $j<N$ $z_{j+N}=\bar z_j$. In the
second sum in (\ref{upoles}) each pair of poles contributed one term. In Eq.(\ref{eqz})
we again employ $2N$ poles since all of them interact. We can write the
pole dynamics in  terms of the real and imaginary parts $x_j$ and $y_j$.
Because of the
arrangement in pairs it is sufficient to write the equation for either
$y_j>0$ or for
$y_j<0$. We opt for the first. The equations for the positions of the poles read
\begin{eqnarray}
&&-L^2{dx_{j}\over dt}=\nu\sum_{k=1,k\neq j}^N
   \sin(x_j-x_k)\Bigg[
   [\cosh (y_j-y_k) \label{xj} \\ && -\cos (x_j-x_k)]^{-1}+[\cosh
(y_j+y_k)-\cos (x_j-x_k)]^{-1}\Bigg]  \nonumber\\
&& L^2{dy_{j}\over dt}=\nu\sum_{k=1,k\neq j}^{N }\Big({\sinh (y_j-y_k)\over
   \cosh (y_j-y_k)-\cos (x_j-x_k)}\nonumber \\ &&+
   {\sinh (y_j+y_k)\over \cosh (y_j+y_k)-\cos (x_j-x_k)}
   \Big)+\nu\coth (y_j)- L\label{yj}  .
\end{eqnarray}
We note that if the initial conditions of the differential equation
(\ref{eqfinal})
are expandable in a finite number of poles, these equations of motion
preserve this number
as a function of time. On the other hand, this may be an unstable situation
for the
partial differential equation, and noise
can change the number of poles. This issue will be examined at length in
Section
\ref{noise}. We turn now to a discussion of the steady state solution of
the equations
of the pole-dynamics.
%%%%%%%%%%%%%%%%%%%%%%%%%%
\subsection{Qualitative properties of the stationary solution}

The steady-state solution of the flame front propagating in channels of
width $2\pi$
was presented in Ref.\cite{85TFH}. Using these results we can immediately
translate
the discussion to a channel of width $L$. The main results are summarized
as follows:
\begin{enumerate}
\item There is only one
stable stationary solution which is geometrically represented by a giant
cusp (or equivalently one finger) and
analytically by $N(L)$ poles which are aligned on one line parallel to the
imaginary
axis. The existence of this solution is made clearer with the following remarks.
\item There exists an attraction between the poles
along the real line. This is obvious from Eq.(\ref{xj}) in which the sign
of $dx_j/dt$ is
always determined by $\sin(x_j-x_k)$. The resulting dynamics merges all the $x$
positions of poles whose $y$-position remains finite.
\item The $y$
positions are distinct, and the poles are aligned above each others in positions
$y_{j-1}<y_j<y_{j+1}$ with the maximal being $y_{N(L)}$. This can be understood
from Eq.(\ref{yj}) in which the interaction is seen to be repulsive at
short ranges,
but changes sign at longer ranges.
\item If one adds
an additional pole to such
a solution, this pole (or another) will be pushed to infinity along the
imaginary axis.	
If the system has less than $N(L)$ poles it is unstable to the addition of
poles,
and any noise will drive the system towards this unique state. The number
$N(L)$ is
\begin{equation}
N(L)= \Big[{1 \over 2}\left( {L \over \nu }+1\right) \Big]\ , \label{NofL}
\end{equation}
where $\Big[ \dots \Big]$ is the integer part. To see this consider
a system with $N$ poles and such
that all the values of $y_j$ satisfy the condition $0< y_j <y_{max}$.
Add now one additional pole whose coordinates are $z_a\equiv (x_a,y_a)$ with
$y_a\gg
y_{max}$. From the equation of motion for $y_a$, (\ref{yj}) we see that
the terms in the sum are all of the order of unity as is also the
$\cot(y_a)$ term. Thus
the equation of motion of $y_a$ is approximately
\begin{equation}
{dy_a \over dt}\approx \nu{2N+1 \over L^2}-{1\over L} \ . \label{ya}
\end{equation}
The fate of this pole depends on the number of other poles.
If $N$ is too large the pole will run to infinity, whereas if $N$ is  small
the pole
will be attracted towards the real axis. The condition for moving away to infinity
is that $N > N(L)$ where $N(L)$ is given by (\ref{NofL}). On the
other hand the $y$ coordinate of the poles cannot hit zero.
Zero is a repulsive line, and poles are pushed away from zero with infinite velocity.
 To see this consider
a pole whose $y_j$ approaches zero. For any finite $L$ the term
$\coth(y_j)$ grows
unboundedly whereas all the other terms in Eq.(\ref{yj}) remain  bounded.
\item  The height of the cusp is proportional to $L$. The distribution of
positions
of the poles along the line of constant $x$ was worked out in \cite{85TFH}.
\end{enumerate}
We will refer to the solution with all these properties as the
Thual-Frisch-Henon
(TFH)-cusp solution.
Next we turn to the stability analysis of this solution.
%%%%%%%%%%%%%%%%%%%%%%%%%%%%%%%%%%%%%%%%
\section{Linear Stability Analysis in Channel Geometry}

In this section we discuss the linear stability of the TFH-cusp solution. To
this aim we first use  Eq.(\ref{upoles}) to write the
steady solution $u_s(\theta)$ in the form:
\begin{equation}
u_s(\theta)=\nu\sum_{j=1}^{N}{2\sin [\theta-x_s]\over
\cosh [y_j]-\cos [\theta-x_s]}\ , \label{stat}
\end{equation}
where $x_s$ is the real (common) position of the stationary poles and $y_j$
their
stationary imaginary position. To study the stability of this solution we need
to determine the actual positions $y_j$. This is done numerically by integrating the
equations of motion for the poles starting from $N$ poles in initial
positions and
waiting for relaxation. Next one perturbs this solution with a
small perturbation $\phi(\theta,t)$: $u(\theta,t) =
u_s(\theta)+\phi(\theta,t)$ .
 Linearizing the dynamics for small $\phi$ results in the equation of motion
\begin{eqnarray}
 &&{\partial \phi(\theta,t) \over \partial t}= {1\over L^2}\Big
[\partial_\theta [u_s(\theta) \phi(\theta,t)]
 +\nu \partial_\theta^2\phi(\theta,t)\Big]\nonumber \\
 && +{1\over L}I(\phi(\theta,t)) \ . \label{linear}
\end {eqnarray}
\subsection{Fourier decomposition and eigenvalues}
The linear equation can be decomposed in Fourier modes according to 
\begin{eqnarray}
\phi(\theta,t)&=&\sum_{k=-\infty}^{\infty} \hat\phi_k(t) e^{ik\theta}\\
u_s(\theta)&=&-2{\nu}i\sum_{k=-\infty}^{\infty}
  \sum_{j=1}^N {\rm sign}(k)e^{-\mid k \mid y_j}e^{ik\theta}
\end {eqnarray}
In these sums the discrete $k$ values run over all the integers.
Substituting in (\ref{linear}) we get the equations:
\begin{equation}
{d\hat \phi_k(t) ) \over dt}= \sum _n a_{kn} \hat\phi_n(t)\ , \
\end {equation}
 where  $a_{kn}$ is a infinite matrix whose entries are given by
\begin{eqnarray}
a_{kk}&=&{\mid k \mid\over L} -{\nu \over L^2} k^2\\
a_{kn}&=&{k\over L^2}{\rm sign}(k-n)({2\nu} \sum_{j=1}^N e^{-\mid k-n \mid y_j})
 \quad k \neq n \ .
\end {eqnarray}
To solve for the eigenvalues of this matrix we need to truncate it at some
cutoff $k$-vector
$k^*$. The choice of $k^*$ can be based on the linear
stability
analysis of the flat front. The scale $k_{max}$, cf. (\ref{kmax}),
is the largest $k$ which is still linearly unstable.
We must choose $k^*>k_{max}$ and test the choice by the converegence of the
eigenvalues.
The chosen value of $k^*$ in our numerics was $4k_{max}$.
The results for the low order eigenvalues of the matrix $a_{kn}$ that were
obtained
from a converged numerical calculation are presented in Fig.1.
The eigenvalues are multiplied by $L^2$ and are plotted
as a function of $L$. We order the eigenvalues
in decreasing order and denote them as $\lambda_0\le \lambda_1\le \lambda_2
\dots$.
The figure offers a number of qualitative observations:
\begin{enumerate}
\item There exists an obvious Goldstone or translational mode $u_s'(\theta)$
with eigenvalue $\lambda_0=0$, which is shown with rhombes in Fig.1.
This eigenmode stems from the Galilean invariance of the equation of motion.
\item The eigenvalues oscillate periodically between values that are
$L$-independent in this
presentation (in which we multiply by $L^2$). In other words, up to the
oscillatory
behavior the eigenvalues depend on $L$ like $L^{-2}$.
\item The eigenvalues $\lambda_1$ and $\lambda_2$, which are represented by squares
and circles in Fig.1, hit zero periodically. The functional
dependence in this presentation appears almost piecewise linear.
\item The higher eigenvalues also exhibit similar qualitative behaviour,
but without
reaching zero.
We note that the solution becomes marginally stable for every value of $L$ for
which the eigenvalues
$\lambda_1$ and $\lambda_2$ hit zero. The $L^{-2}$ dependence of the spectrum
indicates that the solution becomes more and more sensitive to noise as $L$
increases.
\end{enumerate}
\subsection{Qualitative understanding using pole-analysis}

The most interesting qualitative aspects are those enumerated above as item
2 and 3.
To understand them it is useful to return to the pole description, and to
focus on
Eq.(\ref{ya}). This equation describes the dynamics of a single far-away
pole. We
remarked before that this equation shows that for {\em fixed} $L$ the
stable number
of poles is the integer part (\ref{NofL}). Define now the number $\alpha$,
$0\le\alpha\le 1$, according to
\begin{equation}
\alpha= \Big[{1 \over 2}\left( {L \over \nu }+1\right) \Big]-
{1 \over 2}\left( {L \over \nu }-1\right)
 \ . \label{alpha}
\end{equation}
Using this number we rewrite Eq.(\ref{ya}) as
\begin{equation}
{dy_a\over dt} \approx {2\nu \over L^2}\alpha \ . \label{yalpha}
\end{equation}
As $L$ increases, $\alpha$ oscillates piecewise linearly and periodically
between zero and unity. This shows that a distant pole which is added to
the giant
cusp solution is usually repelled to infinity except when $\alpha$ hits
zero and the system becomes marginally unstable to the addition of a new pole. 

To connect this to the linear stability analysis we note from Eq.(\ref{upoles})
that a single far-away pole solution (i.e with $y$ very large) can
be written as
\begin{equation}
u(\theta,t) = 4 \nu e^{-y(t)} \sin(\theta-x(t)) \ . \label{fary}
\end{equation}
Suppose that we add to our giant cusp solution a perturbation of this
functional form.
>From Eq.(\ref{yalpha}) we know that $y$ grows linearly in time, and therefore
this solution decays exponentially in time. The rate of decay is a linear
eigenvalue of the stability problem, and from Eq.(\ref{yalpha}) we understand
both the $1/L^2$ dependence and the periodic marginality. We should
note that this way of thinking gives us a significant part of the $L$ dependence
of the eigenvalues, but not all. The variable $\alpha$ is rising from zero
to unity
periodically, but after reaching unity it hits zero instantly. Accordingly, if
the highest non zero eigenvalue were fully determined by the pole analysis,
we would expect this eigenvalue to behave as the solid line shown in Fig.2.
The actual highest eigenvalue computed from the stability matrix is shown in 
rhombes connected by dotted line.
It is clear that the pole analysis gives us a great deal of qualitative
and quantitative understanding, but not all the features agree.
%%%%%%%%%%%%%%%%%%%%%%%%%%%%%%
\subsection{Dynamics near marginality}

The discovery of marginality at isolated values of $L$ poses questions regarding
the fate of poles that are added at very large $y$'s at certain $x$-positions.
We will argue now that when the system becomes marginally stable, a new pole can
be added to those existing in the giant cusp. We remember that these poles
have a common $\theta$ position that we denote as $\theta=\theta_c$. The fate of
a new pole added at infinity depends on its $\theta$ position. If the position
of the new pole is again denoted as $y_a$, and $\infty \gg y_a \gg y_{max}$, we
can see from Eq.(\ref{yj}) that $dy_a/dt$ is maximal when $\theta_a=\theta_c$,
whereasit is minimal when $\theta_a-\theta_c =\pi$. This follows from the fact that
the cosine term has a value $+1$ when $\theta_a=\theta_c$ and a value $-1$ when
 $\theta_a-\theta_c =\pi$. For large $y$ differences
the terms in the sum take on their minimal value when the $\cos$ term is $-1$
and their maximal values at $+1$. For infinitely large $y_a$ the equation
of motion
is (\ref{ya}) which is independent of $\theta_a$. Since the RHS of this
equation becomes
zero at marginality, we conclude that for very large but finite $y_a$ 
$dy_a/dt$ changes sign from positive to negative when $\theta_a-\theta_c$
changes from
zero to $\pi$. The meaning of this observation is that the most unstable
points in the
system are those points which are furthest away from the giant cusp.
It is interesting to discuss the fate of a pole that is added to the system at
such a position. From the point of view of the pole dynamics 
$\theta=\theta_c+\pi$
is an unstable fixed point for the motion along the  $\theta$ axis. The
attraction to the giant
cusp exactly vanishes at this point. If we start with a pole at a very large
$y_a$ close to this value of $\theta$ the down-fall along the $y$
coordinate will be faster than the lateral motion towards the giant cusp.
We expect
to see therefore the creation of a small cusp at $\theta$ values close to
$\pi$ that
precedes a later stage of motion in which the small cusp moves to
merge with the giant cusp. Upon the approach of the new pole to the giant
cusp all
the existing poles will move up and
the furthest pole at $y_{max}$ will be kicked off to infinity. We will
later explain
that this type of dynamics occurs in stable systems that are driven by noise.
The noise generates far
away poles (in the imaginary direction) that get attracted around
$\theta=\theta_c+\pi$ to
create small cusps that run
continuously towards the giant cusp.
%%%%%%%%%%%%%%%%%%%%%
\subsection{Nonlinear Stability}

The intuition gained so far can be used to discuss the issue of stability of a
stable system to {\em larger} perturbations. In other words, we may want to
add to the
system poles at finite values of $y$ and ask about their fate. We first
show in this
subsection that poles whose initial $y$ value is below $y_{max}\sim \log
(L^2/\nu^2)$
will be attracted towards the real axis. The scenario is
similar to
the one described in the last paragraph.

Suppose that we generate a stable system with a giant cusp at $\theta_c=0$ with
poles distributed along the $y$ axis
up to $y_{max}$. We know that the sum of all the forces that act on the
upper  pole
is zero. Consider then an additional pole inserted in the position $(\pi,y_{max})$.
It is obvious from Eq.(\ref{yj}) that the forces acting on this pole will
pull it downward. On the other hand if its initial position is much above
$y_{max}$
the force on it will be repulsive towards infinity. We see that this simple
argument identifies $y_{max}$ as the typical scale for nonlinear instability.

Next we estimate $y_{max}$ and interpret our result in terms of the {\em
amplitude}
of a perturbation of the flame front. We explained that uppermost pole's 
position fluctuates between a minimal value and infinity as $L$ is changing.
We want to estimate the characteristic scale of the minimal value of $y_{max}(L)$.
To this aim we employ the result of ref.\cite{85TFH}
regarding the stable distribution of pole positions in a stable large
system. The
parametrization of \cite{85TFH} differs from ours; to go from our
parametrization in
Eq.(\ref{eqfinal}) to theirs we need to rescale $u$ by $L^{-1}$ and $t$ by
$L$. The parameter $\nu$ in their parameterization is $\nu/L$ in ours.
According to \cite{85TFH} the number of poles between $y$ and $y+dy$ is given
by the $\rho(y)dy$ where the density $\rho(y)$ is
\begin{equation}
\rho(y)={L\over \pi^2\nu}\ln[\coth(|y|/4)] \ . \label{dist}
\end{equation}
To estimate the minimal value of $y_{max}$ we require that the tail of the
distribution $\rho(y)$ integrated between this value and infinity will allow
one single pole. In other words, 
\begin{equation}
\int_{y_{max}}^\infty dy \rho(y) \approx 1 \ . \label{integ}
\end{equation}
Expanding (\ref{dist}) for large $y$ and integrating explicitly the result
in (\ref{integ})
we end up with the estimate
\begin{equation}
y_{max} \approx 2\ln\Big[{4L\over \pi^2\nu}\Big]
\end{equation}
For large $L$ this result is $y_{max}\approx \ln({L^2 \over \nu^2})$. If we now add an
additional
pole in the position $(\theta,y_{max})$ this is equivalent to perturbing the solution $u(\theta,t)$
with a function $\nu e^{-y_{max}} \sin(\theta)$, as can be seen directly
from (\ref{upoles}).
We thus conclude that the system is unstable to a perturbation {\em larger} than
\begin{equation}
u(\theta) \sim \nu^3 \sin(\theta)/L^2 \ . \label{nu3L2}
\end{equation}
This indicates a very strong size dependence of the sensitivity of the giant
cusp solution to external perturbations. This will be an important ingredient in our
discussion of noisy systems.
%%%%%%%%%%%%%%%%%%%%%%%%%%%%%%%%%%%%%%
\section{Initial Conditions, Pole Decomposition and Coarsening}

In this section we show first that any initial conditions can be approximated
by pole decomposition. Later, we show that the dynamics of sufficiently
smooth initial
data can be well understood from the pole decomposition. Finally we employ this
picture to describe the {\it inverse cascade} of cusps into the giant cusp which is
the final steady state.  By inverse cascade we mean a nonlinear coarsening process in which
the small scales coalesce in favor of larger scales and finally the system
staturates at the largest available scale\cite {coars}.  
%%%%%%%%%%%%%%%%%%%%
\subsection{Pole Expansion: General Comments}
The fundamental question is how many poles are needed to
describe any given initial condition. The answer, of course, depends on how
smooth are
the initial conditions. Suppose also that we have an initial function
$u(\theta,t=0)$ that is $2\pi$-periodic and which at time $t=0$ admits a Fourier
representation
\begin{equation}
u(\theta) = \sum_{k=1}^{\infty} A_k \sin{(k\theta+\phi_k)} \ , \label{Fourier}
\end{equation}
with $A_k > 0$ for all $k$. Suppose that  we want to find a
pole-decomposition representation $u_p(\theta)$ such that
\begin{equation}
|u_p(\theta) - u(\theta)| \le \epsilon	\qquad {\rm for~~every~~}\theta \ ,
\label{eps}
\end{equation}
where $\epsilon$ is a given wanted accuracy.  If $u(\theta)$ is
differentiable we can
cut the Fourier expansion at some finite $k=K$ knowing that the remainder
is smaller
than, say, $\epsilon/2$.  Choose now a large number $M$ and a small number
$\Delta\ll 1/M$
and write the pole representation for $u_p(\theta)$ as
\begin{equation}
u_p(\theta) = \sum_{k=1}^K \sum_{p=0}^{M-1} {2k \sin{(k\theta+\phi_k})\over
\cosh{[k(y_k
+p\Delta)]}-cos{(k\theta+\phi_k})} \ . \label{up}
\end{equation}
To see that this representation is a particular form of the general formula
(\ref{upoles})
We use the following two identities
\begin{equation}
\sum_{k=0}^{\infty} e^{-kt} \sin{xk} = {1\over 2} {\sin{x}\over
\cosh{t}-\cos{x}}
\ , \label{identA}
\end{equation}
\begin{equation}
\sum_{k=0}^{K-1}\sin (x+ky)=\sin (x+{K-1 \over 2}y)\sin{Ky \over 2}
     {\rm cosec}{y \over 2}\ . \label{identB}
\end{equation}
From these follows a third identity
\begin{eqnarray}
&&\sum_{j=0}^{K-1}{2\sin{(x-{2\pi j\over K}+\phi)}\over\cosh{y}-\cos{(x-{2\pi
j\over K}+\phi)}}\nonumber \\
&&= {2K\sin{(Kx+\phi)}\over \cosh{Ky}-\cos{(Kx+\phi)}} \ . \label{identC}
\end{eqnarray}
Note that the LHS of (\ref{identC}) is of the form (\ref{upoles}) with $K$
poles whose
positions are all on the line $y_j=y$ and whose $x_j$ are on the lattice points
$2\pi j/K-\phi$. On the other hand every term in (\ref{up}) is of this form.

Next we use (\ref{identA})
to rewrite (\ref{up}) in the form
\begin{equation}
u_p(\theta) = \sum_{k=1}^K \sum_{p=0}^{M-1}\sum_{n=1}^{\infty} 4k
e^{-nk(y_k+p\Delta)}
\sin{(nk\theta+n\phi_k)} \ . \label{3sum}
\end{equation}
Exchanging order of summation between $n$ and $p$ we can perform the
geometric sum
on $p$. Denoting
\begin{equation}
b_{n,k}\equiv \sum_{p=0}^{M-1} e^{-nkp\Delta}={1-e^{-Mkn\Delta}\over
1-e^{-kn\Delta}}
\ , \label{b}
\end{equation}
we find
\begin{eqnarray}
u_p(\theta) &=& \sum_{k=1}^K \sum_{n=1}^{\infty} 4kb_{n,k} e^{-nky_k}
\sin{(nk\theta+n\phi_k)}  \nonumber \\
&=& \sum_{k=1}^K \sum_{n=2}^{\infty} 4kb_{n,k} e^{-nky_k}
\sin{(nk\theta+n\phi_k)} \nonumber \\ &+&\sum_{k=1}^K  4kb_{1,k} e^{-ky_k}
\sin{(k\theta+\phi_k)}
\ . \label{2sum}
\end{eqnarray}
Compare now the second term on the RHS of (\ref{2sum}) with (\ref{Fourier}).
We can identify
\begin{equation}
e^{-ky_k} = {A_k\over 4kb_{1,k}} \label{choosey}
\end{equation}
The first term can be then bound from above as
\begin{eqnarray}
&&\Big | \sum_{k=1}^K \sum_{n=2}^{\infty} 4kb_{n,k} e^{-nky_k}
\sin{(nk\theta+n\phi_k)}\Big | \label{in1} \\ &&\le  \sum_{k=1}^K 
\sum_{n=2}^{\infty} \Big
|4kb_{n,k}
\left[{A_k\over 4kb_{1,k}}\right]^n \sin{(nk\theta+n\phi_k)\Big|} \nonumber . 
\end{eqnarray}
The sine function and the factor $(4K)^{1-n}$ can be replaced by unity and
we can bound
the RHS of (\ref{in1}) by
\begin{equation}
\sum_{k=1}^K \sum_{n=2}^{\infty}\left[{A_k\over b_{1,k}}\right]^n b_{n,k} 
\le \sum_{k=1}^K A_k \sum_{n=1}^{\infty}\left[{A_k\over b_{1,k}}\right]^n\label{in2} ,
\end{equation}
where we have used the fact that $b_{n,k}\le b_{1,k}$ which follows
directly from (\ref{b}).
Using now the facts that $b_{1,K}\le b_{1,k}$ for every $k\le K$ and
that $A_k$ is
bounded by some finite $C$ since it is a Fourier coefficient, we can bound
(\ref{in2})
by $C^2K /(b_{1,K}-C)$. Since we can select the free parameters $\Delta$
and $M$ to make
$b_{1,K}$ as large as we want, we can make the remainder series smaller in
absolute value
than $\epsilon/2$.

The conclusion of this demonstration is that any initial condition that can
be represented
in Fourier series can be approximated to a desired accuracy by
pole-decomposition. The number
of needed poles is of the order $K^2\times M$. Of course, the number of poles
thus generated
by the initial conditions may exceed the number $N(L)$ found in
Eq.(\ref{NofL}). In such
a case the excess poles will move to infinity and will become irrelevant
for the short
time dynamics. Thus a smaller number of
poles may be needed to describe the state at larger times than at $t=0$.
 We need to stress at this point that the pole
decomposition is
over complete; for example, if there is exactly one pole at $t=0$ and we use the
above technique to reach a pole decomposition we would get a large number
of poles in our representation.
%%%%%%%%%%%%%%%%%%%%%%%%%%%%%%%%%%%%%%%%%%%%%
\subsection{The initial stages of the front evolution: the exponential stage and 
the inverse cascade}

In this section we employ the connection between Fourier expansion and pole
decomposition to understand the initial exponential stage of the evolution of the flame
front with small initial data $u(\theta,t=0)$. Next we employ our knowledge of the pole 
interactions  to explain the slow dynamics of coarsening  into the steady state solution.

Suppose that initially the
expansion (\ref{Fourier}) is available with all the coefficients $A_k\ll 1$.
We know from the linear instability of the flat flame front that each
Fourier component changes exponentially in time according to the linear
spectrum (\ref{spec}). The components with wave vector larger than (\ref{kmax})
decrease, whereas those with lower wave vectors increase. The fastest
growing mode
is $k_c=L/2\nu$. In the linear stage of growth this mode will dominate the
shape of the flame front, i.e.
\begin{equation}
u(\theta,t)\approx A_{k_c}e^{\omega_{k_c}t} \sin(k_c\theta) \ . \label{initial}
\end{equation}
Using Eq.(\ref{identC}) for a large value of $y$ (which is equivalent to small
$A_{k_c}$) we see that to the order of $O(A^2_{k_c})$ (\ref{initial}) can
be represented as a sum over $L/2\nu$ poles arranged periodically along the
$\theta$ axis. Other unstable modes will contribute 
similar arrays of poles but at much higher
values of $y$, since their amplitude is exponentially smaller. In addition we
have nonlinear corrections to the identification of the modes in terms of
poles. These corrections can be again expanded in terms of Fourier modes, and
again identified with poles, which  will  be further away
along the $y$ axis, and with higher frequencies. To see this one can use 
Eq.(\ref{2sum}), subtract from $u_p(\theta)$ the leading pole representations, and
reexpand in Fourier series. Then we identify the leading order with 
double the number of poles that
are situated twice further away along the $y$ axis. 

We note that even when
all the unstable modes are present, the number of poles in the first order
identification 
is finite for finite $L$, since there are only $L/\nu$ unstable modes.
Counting the
number of poles that each mode introduces we get a total number of
$\Big(L/\nu \Big)^2$ poles.
The number $L/2\nu$ of poles which are associated with the most unstable mode
is precisely the number allowed in the stable stationary solution,
cf.(\ref{NofL}).
When the poles approach the real axis and cusps begin to develop, the linear
analysis no longer holds, but the pole description does.

We  now describe the qualitative scenario for
the establishment of the steady state. Firstly, we understand that all the
poles that belong to less unstable
modes will be pushed towards infinity. To see this think of the system at
this stage
as an array of uncoupled systems with  a scale of the order of unity. Each such
system will have a characteristic value of $y$. As we discussed
before poles
that are further away along the $y$ axis will be pushed to infinity.
Therefore the system will remain with the $L/2\nu$ poles of the most unstable mode.
The net
effect of the poles belonging to the (nonlinearly) stable modes is to
destroy the otherwise
perfect periodicity of the poles of the unstable mode. The see the effect of the
higher order correction to the pole identification we again recall that they
can be represented as further away poles with higher frequencies, whose dynamics is 
similar to the less unstable modes that were just discussed. They do not become more
relevant when time goes on.

Once the poles of the
stable modes get sufficiently far from the real axis, the
dynamics of the  remaining poles will begin to develop according to the interactions
that are directed along the real axis.
These interactions are much weaker and the resulting dynamics occur on much longer time
scales.
The qualitative picture is of an inverse cascade of merging the $\theta$
positions of the poles. 
We  note that the system has a set of unstable fixed points which are 'cellular solutions' 
 described by a periodic arrangement of 
poles along the real axis with a frequency $k$.  These fixed points are not stable 
and they collapse,
under perturbations, with a characteristic time scale (that depends on $k$) 
to  the next unstable  fixed point 
at $k'=k/2$. This process then goes on indefinitely until 
$k \sim1/L$ i.e. we reach the  giant cusp, the steady-state stable solution \cite {coars}.  

This scenario is seen very clearly in the numerical simulations. In Fig.3 we show
the time evolution of the flame front starting from small white-noise initial
conditions. The bottom curve pertains to the earliest time in this picture,
just after
the fast exponential growth, and one sees clearly the periodic array of
cusps that form. The successive images show the progress of the flame front
in time, and
one observes the development of larger scales with deeper cusps that
represent the
partial coalescence of poles onto the same $\theta$ positions. In Fig.4 we
show the
width and the velocity of this front as a function of
time. One recognizes the exponential stage of growth in which the $L/2\nu$
poles
approach the $\theta$ axis, and then a clear cross-over to much slower dynamics
in which the effective scale in the system grows with a slower rate. 

The slow dynamics stage 
can be understood qualitatively using the previous interpretation of the 
cascade as follows: if the initial number of poles belonging
to the unstable mode is $L/2\nu$, the initial effective linear scale is
$2\nu$. Thus the
first step of the inverse cascade will be completed in a time scale of the order
of $2\nu$. At this point the effective linear scale doubles to $4\nu$, and
the second
step will be completed after such a time scale. We want to know what is the
typical
length scale $l_t$ seen in the system at time $t$. The typical width of the system at this
stage will be proportional to this scale.

Denote the number of cascade steps that took place
until this
scale is achieved by $s_l$. The total time elapsed, $t(l_t)$ is the sum
\begin{equation}
t(l_t) \sim \sum_{i=1}^{s_l} 2^i \ . \label{tLt}
\end{equation}
The geometric sum is dominated by the largest term and we therefore estimate
$t(l_t)\sim l_t$. We conclude
that the  scale and the width   are  linear in the time elapsed from the initial conditions
($l_t\sim t^\zeta,~\zeta=1$). In noiseless simulations we find (see Fig.4)
a value of $\zeta$ which is $\zeta\approx 0.95 \pm 0.1$.
%%%%%%%%%%%%%%
\subsection{Inverse cascade in the presence of noise\label{ICPN}}
An interesting consequence of the discussion in the last section is that the
inverse cascade process is an effective ``clock" that measures the typical
time scales in this system. For future purposes we need to know the typical
time scales when the dynamics is perturbed by random noise. To this aim we
ran simulations following the inverse cascade in the {\it presence} of external
noise. The main result that will be used in later arguments is that now the
appearance of a typical scale $l_t$ occurs not after time $t$, but rather
according to
\begin{equation}
l_t\sim t^{\zeta}\ , \quad \zeta\approx 1.2\pm 0.1\ . \label{timescale}
\end{equation}
The numerical confirmation of this law is exhibited in Fig.5 . We also find
that the front velocity in this case increases with time according to
\begin{equation}
v \sim t^\gamma \ , \quad \gamma\approx 0.48\pm 0.05\ . \label{defgamma}
\end{equation}
This result will be related to the acceleration of the flame front in noisy
simulations, as will be seen in the next Sections. The result
(\ref{timescale}) will
be helpful in Section \ref{accel} in estimating the values of the scaling
exponents.
%%%%%%%%%%%%%%%%%%%%%%%%%%%%%%%%%%%%%%%%%%
\section{Acceleration of the Flame Front, Pole Dynamics and Noise}
\label{noise}
A major motivation of this Section is the observation that in radial
geometry the 
same equation of motion shows an acceleration of the flame
front. The aim of this section is to argue that this phenomenon is caused by the
noisy generation of new poles. Moreover, it is our contention that a great
deal can
be learned about the acceleration in radial geometry by considering the
effect of
noise in channel growth. In Ref. \cite{85TFH} it was shown that any initial
condition
which is represented in poles
goes to a unique stationary state which is the giant cusp which propagates with
a constant velocity $v=1/2$ up to small $1/L$ corrections. In light of our
discussion
of the last section we expect that any smooth enough initial condition will
go to
the same stationary state. Thus if there is no noise in the dynamics of a finite
channel, no acceleration of the flame front is possible. What happens if we
add noise to the system?

For concreteness we introduce an additive white-noise term $\eta(\theta,t)$to
the equation of motion  (\ref{eqfinal})  where
\begin{equation}
\eta(\theta,t) = \sum_k{\eta_k(t) \exp{(ik\theta)}}\ , \label{eta}
\end{equation}
and the Fourier amplitudes $\eta_k$ are correlated according to
\begin{equation}
<\eta_k(t)\eta^{*}_{k'}(t')>=f\delta(k-k')\delta(t-t') \ . \label{corr}
\end{equation}
We will first examine the result of numerical simulations of noise-driven
dynamics,
and later return to the theoretical analysis.
%%%%%%%%%%%%%%%%%%%%%%
\subsection{Noisy Simulations}

Previous numerical  investigations \cite{94FSF,90GS} did not introduce noise in
a controlled
fashion. We will argue later that some of the phenomena encountered in
these simulations
can be ascribed to the (uncontrolled) numerical noise. We performed numerical
simulations of Eq.(\ref{eqfinal} using a pseudo-spectral method. The time-stepping scheme
was chosen as Adams-Bashforth with 2nd order presicion in time. The additive white noise was
generated in Fourier-space by choosing
$\eta_k$
for every $k$ from a flat distribution
in the interval $[-\sqrt{2f},\sqrt{2f}]$. We examined the average steady state
velocity of the front
as a function of $L$ for fixed $f$ and as a function of $f$ for fixed $L$.
We found the
interesting phenomena that are summarized here:
\begin{enumerate}
\item When the noise level $f$ is fixed  the average velocity $v$
increases with
$L$, see Fig.6. There are two different regimes of this behavior. For
sufficiently
small values of $f$ and $L$ we observe a power law dependence of $v$ on $L$:
\begin{equation}
v\sim L^\mu , \quad \mu\approx 0.42\pm 0.03 \ . \label{scale1}
\end{equation}
For large values of $f$ and $L$ the dependence of $v$ on $L$ is much
stronger, $\mu >1.5$, but we cannot estimate it well because of lack of dynamical range.
 For
a given value of $f$ the transition between the two regimes appears upon
increasing
$L$, and we denote the critical value as $L_c(f)$, see Fig.7. The data
indicates that
\begin{equation}
L_c(f)\sim f^{-\alpha} \ , \label{Lcfn}
\end{equation}
with $\alpha\approx 1.0\pm 0.2$.
\item When the system size $L$ is fixed the average velocity depends on $f$ as
\begin{equation}
v\sim f^\xi \ . \label{vf}
\end{equation}
For sufficiently small value of $f$ this dependence is very weak, and
$\xi\approx 0.02$,
see Fig.8. For
large values of $f$ the dependence is much stronger, $\xi=1\pm 0.1$.
\item The locus separating weaker dependencies of $v$ on $L$ and $f$
(regime I) from stronger
dependencies (regime II) is shown in Fig.7. The giant cusp remains
recognizable in
both regimes. In regime I one observes a series of small cusps superposed on the
structure of the giant cusp. In   Fig.9 we show a set of interfaces with a growing size and 
with the same noise.  One can observe 
a strong increase in the number of cusps and the complexity of their arrangements
as the system size increases.
In regime II there are strong
fluctuations in the gradient field $u(\theta,t)$, see Fig.10.
\item Measurements of the width of the front indicate that it has a a very weak
dependence on $f$ for fixed $L$ in regime I.
\end{enumerate}
%%%%%%%%%%%%%%%%%%%%%%%%%%%%%%%%%%%%
\subsection{Theoretical Discussion of the Effect of Noise}
\subsubsection{The Threshold of Instability to Added Noise}
\label {regime0}
First we present the theoretical arguments that explain the sensitivity of the
giant cusp solution to the effect of added noise. This sensitivity increases
dramatically with increasing the system size $L$. To see this we use again the
relationship between the linear stability analysis and the pole dynamics.

Our additive noise introduces perturbations with all $k$-vectors. We showed
previously
that the most unstable mode is the $k=1$ component $A_1 \sin(\theta)$. Thus the
most effective noisy perturbation is $\eta_1 \sin(\theta)$ which can
potentially lead
to a growth of the most unstable mode. Whether or not this mode will grow
depends
on the amplitude of the noise. To see this clearly we return to the pole
description.
For small values of the amplitude $A_1$ we represent $A_1 \sin(\theta)$ as a
single
pole solution of the functional form $\nu e^{-y}\sin{\theta}$. The $y$
position is
determined from $y=-\log{|A_1| /\nu}$, and the $\theta$-position is
$\theta=\pi$ for
positive $A_1$ and $\theta=0$ for negative $A_1$. From the analysis of
Section III we know that
for very small $A_1$ the fate of the pole is to be pushed to infinity,
independently
of its $\theta$ position; the dynamics is symmetric in $A_1\to -A_1$ when $y$ is
large enough. On the other hand when the value of $A_1$ increases the symmetry
is broken and the $\theta$ position and
the sign of $A_1$ become very important. If $A_1>0$ there is a threshold value
of $y$ below which the
pole is attracted down. On the other hand if $A_1<0$, and $\theta=0$ the
repulsion
from the poles of the giant cusp grows with decreasing $y$. We thus
understand that
qualitatively speaking the dynamics of $A_1$ is characterized by an asymmetric
``potential" according to
\begin{eqnarray}
\dot A_1 &=& -{\partial V(A_1)\over \partial A_1}\ , \label{dvda}\\
V(A_1) &=& \lambda A_1^2 -aA_1^3+\dots \ . \label{poten}
\end{eqnarray}
>From the linear stability analysis we know that $\lambda\approx \nu/L^2$,
cf. Eq.(\ref{ya}).
We know further that the threshold for nonlinear instability is at
$A_1\approx \nu^3/L^2$,
cf. Eq(\ref{nu3L2}). This determines that value of the coefficient
$a\approx 2/3\nu^2$. The
magnitude of the ``potential" at the maximum is
\begin{equation}
V(A_{max}) \approx \nu^7/L^6 \ . \label{vmax}
\end{equation}
The effect of the noise on the development of the mode $A_1\sin{\theta}$ can be
understood from the following stochastic equation
\begin{equation}
\dot A_1 = -{\partial V(A_1)\over \partial A_1}+\eta_1(t) \ . \label{stochA}
\end{equation}
It is well known \cite{Ris} that for such dynamics the rate of escape $R$
over the ``potential" barrier
for small noise is proportional to
\begin{equation}
R\sim {\nu\over L^2} \exp^{-\nu^7/fL^6} \ . \label{wow}
\end{equation}
The conclusion is that any arbitrarily tiny noise becomes effective when the
system size increase and when $\nu$ decreases. If we drive the system with noise
of amplitude $f$ the system can always be sensitive to this noise when its size
exceeds a critical value $L_c$ that is determined by $f \sim \nu^7/L_c^6$.
For $L>L_c$
the noise will introduce new poles into the system.
Even numerical noise in simulations involving large size systems may have
a macroscopic influence.

The appearance of new poles must increase the velocity of the front. The
velocity
is proportional to the mean of $(u/L)^2$. New poles distort the giant cusp by
additional smaller cusps on the wings of the giant cusp, increasing $u^2$. Upon
increasing the noise amplitude more and more smaller cusps appear in the front,
and inevitably the velocity increases. This phenomenon is discussed
quantitatively in
Section \ref{noise}.
%%%%%%%%%%%%%%%%%%
\subsubsection{The Noisy Steady State and its Collapse with
Large Noise and System Size}
In this subsection we discuss the response of the giant cusp solution to
noise levels that are able to introduce a large number of excess poles in
addition
to those existing in the giant cusp. We will denote the excess number of poles
by $\delta N$. The first question that we address is how difficult is it to
insert yet an additional pole when there is already a given excess $\delta
N$. To
this aim we estimate the effective potential $V_{\delta N}(A_1)$ which is
similar to
(\ref{poten}) but is taking into account the existence of an excess number
of poles.
A basic approximation that we employ is that the fundamental form of the
giant cusp solution is not seriously modified by the existence of an excess
number 
of poles. Of course this approximation breaks down quantitatively already 
with one excess pole. 
Qualitatively however it holds well until the excess number of
poles
is of the order of the original number $N(L)$ of the giant cusp solution.
Another approximation is that the rest of the linear modes play no role in this case.
At this
point we limit the discussion therefore to the situation $\delta N\ll N(L)$.

To estimate the parameter $\lambda$ in the effective potential we consider the
dynamics of one pole whose $y$ position $y_a$ is far above $y_{max}$. According
to Eq.(\ref{ya}) the dynamics reads
\begin{equation}
{dy_a\over dt}\approx {2\nu (N(L)+\delta N)\over L^2} -{1\over L}
\end{equation}
Since the $N(L)$ term cancels against the $L^{-1}$ term (cf. Sec. II A), 
we remain
with a repulsive term that in the effective potential translates to
\begin{equation}
\lambda={\nu\delta N\over L^2} \ . \label{lambda2}
\end{equation}
Next we estimate the value of the potential at the break-even point between
attraction
and repulsion. In the last subsection we saw that a foreign pole has to be
inserted below $y_{max}$
in order to be attracted towards the real axis. Now we need to push the new pole
below the position of the existing  pole whose index is $N(L)-\delta N$. This
position is estimated as in Sec III C by employing the TFH distribution
function (\ref{dist}).
We find
\begin{equation}
y_{\delta N}\approx 2\ln{\Big[{4L\over \pi^2\nu\delta N}\Big]} \ . \label{ydelN}
\end{equation}
As before, this implies a threshold value of the amplitude of single pole
solution
$A_{max}\sin{\theta}$ which is obtained from equating $A_{max}=\nu
e^{y_{\delta N}}$.
We thus find in the present case $A_{max}\sim \nu^3(\delta N)^2/L^2$. Using
again
a cubic representation for the effective potential we find $a=2/(3\nu^2\delta
N)$ and
\begin{equation}
V(A_{max}) = {1\over 3}{\nu^7(\delta N)^5\over L^6}\ . \label{max2}
\end{equation}
Repeating the calculation of the escape rate over the potential barrier we find
in the present case
\begin{equation}
R\sim {\nu\delta N\over L^2} \exp^{-\nu^7(\delta N)^5/fL^6} \ . \label{wow2}
\end{equation}

For a given noise amplitude $f$ there is always a value of $L$ and $\nu$ for
which
the escape rate is of $O(1)$ as long as $\delta N$ is not too large. When
$\delta N$
increases the escape rate decreases, and eventually no additional poles can
creep
into the system. The typical number $\delta N$ for fixed values of the
parameters is
estimated from equating the argument in the exponent to unity:
\begin{equation}
\delta N\approx \left(fL^6/\nu^7\right)^{1/5} \ . \label{deltaN}
\end{equation}
The most important consequence of this relation is that $\delta N$
increases with
$L$ faster than $N(L)$. Accordingly we expect a breakdown of this picture and
of the
weak noise behavior when $\delta N\approx N(L)$, which is occurring when
$L$ reaches
a critical value $L_c(f)$ where
\begin{equation}
L_c(f) \sim f^{-1} \ . \label{Lcft}
\end{equation}
This prediction is in good quantitative agreement with
(\ref{Lcfn}) supporting the analytical theory. 
%%%%%%%%%%%%%%%%%%%%%%%%%%%%%%%%%%
\subsection{The acceleration of the flame front due to noise}
\label{accel}
In this section we estimate the scaling exponents that characterize the velocity
of the flame front as a function of the system size. Our arguments in this
section
are even less solid than the previous ones, but nevertheless we believe that we succeed
to capture
some of the essential qualitative physics that underlies the interaction between
noise and instability and which results in the acceleration of the flame front.

To estimate the velocity of the flame front we need to write down an
equation for the
mean of $<dh/dt>$ given an arbitrary number $N$ of poles in the system.
This equation follows directly from (\ref{Eqdim}):
\begin{equation}
\left<{dh\over dt}\right>={1  \over L^2}{1 \over
2\pi}\int_{0}^{2\pi}u^2d\theta  \ .
\label{eqr0}
\end{equation}
After substitution of (\ref{upoles}) in (\ref{eqr0}) we get, using (\ref{xj})
and (\ref{yj})
\begin{equation}
\left<{dh\over dt}\right>=2\nu\sum_{k=1}^N {dy_k\over dt}+2\left( {\nu N\over
     L}-{\nu^2 N^2\over L^2}\right)  \ . \label{r0pole}
\end{equation}
The estimates of the second and third terms in this equation are
straightforward.
Writing $N=N(L)+\delta N(L)$ and remembering that $N(L)=\nu/L$ we find that
these terms contribute $-\nu \delta N(L)/L \sim -L^{1/5}$. The first term
will contribute
only when the current of poles is asymmetric. Since noise introduces poles
at a finite
value of $y$, whereas the rejected poles stream towards infinity, we have
an asymmetry that contributes to the velocity of the front. To estimate the first
term we remind the reader of our discussion in Section \ref {ICPN}. In this problem the typical 
time-scale for the poles is the coalescense time of poles with an initial distance 
$L$ in the $x$ direction. 
In noiseless conditions the typical time scales like $L$.
In the presence of noise (cf. Eq.(\ref{timescale})) we found numerically in \ref {ICPN}
that it scales  like $L^{1\over\zeta}$.
Accordingly, the typical flux of poles can be estimated as $\delta
N(L)/L^{1\over \zeta}$.
Thus the current
$\sum \dot y_k$ has a stronger dependence on $L$, i.e $L^{6/5-1/\zeta}$.
Taking the
numerical value of $\zeta=1.2$ we
conclude that Eq.(\ref{r0pole}) predicts a scaling law (\ref{scale1}) with
$\mu=0.37$, in reasonable agreement with the numerics.

We should stress at this point that the argument is not complete. Firstly, we used
the inverse cascade measurement to invoke a typical time scale for the coalescence
of poles by motion along the $x$ axis when the distance between them is $L$, and we
used this time scale for the coalescence of poles in a system whose integral scale is
$L$. This can be only taken as a lower bound the exponent characterizing the time scale,
beacuse of the intervention of additional modes in the larger system. The simple 
identification is a sort of ``single-mode" approximation in which the dynamics is
carried by the most unstable mode only. Secondly, we measured the
exponent relating the velocity to the amplitude of external noise, cf.
Eq.(\ref{vf})
and found that $\xi$ is considerably smaller than the value $1/5$ which is
predicted by
the previous argument. This indicates that the typical time scale also has
an explicit $f$ dependence which is physically plausible but it was not
measured in our simulations. \\~\\
%%%%%%%%%%%%%%%%%%%%%%%%%%%%%%%%%%%%%%%%%%%%%%%%%%%%
\section{summary and conclusions}

The main two messages of this paper are: (i) There is an important
interaction between
the instability of developing fronts and random noise; (ii) This
interaction and its
implications can be understood qualitatively and sometimes quantitatively using
the description in terms of complex poles.

The pole description is natural in this context firstly because it
provides an
exact (and effective) representation of the steady state without noise.
Once one succeeds
to describe also the {\em perturbations} about this steady state in terms
of poles,
one achieves a particularly transparent language for the study of the
interplay between
noise and instability. This language also allows us to describe in
qualitative and
semi-quantitative terms the inverse cascade process of increasing typical
lengths when
the system relaxes to the steady state from small, random initial conditions.

The main conceptual steps in this paper are as follows: firstly one
realizes that
the steady state solution, which is characterized by $N(L)$ poles aligned along
the imaginary axis is marginally stable against noise in a periodic array of
$L$ values. For all values of $L$ the steady state is nonlinearly unstable
against noise. The main and foremost effect of noise of a given amplitude $f$ is 
to introduce
an excess number of poles $\delta N(L,f)$ into the system. The existence of this
excess number of poles is responsible for the additional wrinkling of the
flame front on top of the giant cusp, and for the observed acceleration of
the flame front.
By considering the noisy appearance of new poles we rationalize the
observed scaling laws as a function of the noise amplitude and the system size.

Theoretically we therefore concentrate on estimating $\delta N(L,f)$. The
measurements
do not test our theoretical consideration directly, but rather test the
dependence of
the velocity on $L$ and $f$. The only direct test for our theory is the critical line
shown in Fig.7. The measured exponent is in accord with our analytic
estimates.   
Neverthless we note  that some of our consideration are only qualitative. For example,
we estimated $\delta N(L,f)$ by assuming that the giant cusp solution is not
seriously perturbed. On the other hand we find a flux of poles going to
infinity due
to the introduction of poles at finite values of $y$ by the noise. The
existence of poles
spread between $y_{max}$ and infinity {\em is} a significant perturbation of the
giant cusp solution. Thus also the
comparison between the various scaling exponents measured and predicted must be done
with caution; we cannot guarantee that those cases in which our prediction
hit close
to the measurement mean that the theory is quantitative. However we
believe that
our consideration extract the essential ingredients of a correct theory.

The ``phase diagram" as a function of $L$ and $f$ in this system consists of
three regimes. In the first one, discussed in 
Section \ref {regime0} , the noise is
too small
to have any effect on the giant cusp solution. In the second the noise
introduces excess
poles that serve to decorate the giant cusp with side cusps. In this regime
we find
scaling laws for the velocity as a function of $L$ and $f$ and we are reasonably
successful in understanding the scaling exponents. In the third regime the
noise is large enough to create small scale structures that are not neatly
understood
in terms of individual poles. It appears from our numerics that in this
regime the
roughening of the flame front gains a contribution from the the small scale
structure in a way that is reminiscent of {\em stable}, noise driven growth
models
like the Kardar-Parisi-Zhang model.

One of our main motivations in this research was to understand the
phenomena observed
in radial geometry with expanding flame fronts. A full analysis of this problem
cannot be presented here. We note however that many of the insights offered
above
translate immediately to that problem. Indeed, in radial geometry the flame
front
accelerates and cusps multiply and form a hierarchic structure as time
progresses.
Since the radius (and the typical scale) increase in this system all the
time, new
poles will be added to the system even by a vanishingly small noise. The
marginal
stability found above holds also in this case, and the system
will allow the introduction of excess poles as a result of noise. The
results discussed
in Ref.\cite{96KOP} can be combined with the present insights to provide a theory
of radial growth. This theory will be offered in a forthcoming publication.

Finally, the success of this approach in the case of flame propagation
raises hope
that Laplacian growth patterns may be dealt with using similar ideas. A problem
of immediate interest is Laplacian growth in channels, in which a finger
steady-state solution is known to exist. It is documented that the stability
of such a finger solution to noise decreases rapidly with increasing the channel
width. In addition, it is understood that noise brings about additional
geometric
features on top of the finger. There are enough similarities here to indicate
that a careful analysis of the analytic theory may shed as much light on that
problem as on the present one.

\noindent
{\bf Acknowledgments} This work was supported in part by the German
Israeli Foundation, the US-Israel Binational Science Foundation,
the Minerva Center for Nonlinear Physics, and the
Naftali and Anna Backenroth-Bronicki Fund for Research in Chaos and
Complexity. \\~\\
{\Large \bf Figures Legends}\\~\\
Fig.1: The first 10 highest eigenvalues of the stability matrix with $\nu=1$, 
multiplied by
the square of the system size $L^2$ vs. the system size $L$. Note that all
the eigennvalues
oscillate around fixed values in this presentation, and that the highest
two eigenvalues hit zero periodically. \\~\\
Fig.2: Comparison of the numerically determined highest 4 eigenvalues of
the stability
matrix with the prediction of the pole analysis. The eigenvalues of the stability
matrix are :$\lambda_0$ (squares), $\lambda_1$ (rhombi), $\lambda_2$ (triangles) and
$\lambda_3$ (slanted triangles). The pole analysis (solid line) provides a
qualitative understanding of the stability, and appears to overlap with the highest
eigenvector over half of the range, and with the fourth eigenvalue 
over the other half.\\~\\
Fig.3: The inverse cascade process of coarsening that occurs after preparing the
system with random, small initial conditions. One sees that at successive times
the typical scale
increases until the giant cusp forms, and attracts all the other side-poles. The
effect of the existing numerical additive noise is to introduce poles
that appear as side cusps that are continuously attracted to the giant
cusp. This
effect is obvious to the eye only after the typical scale is sufficiently
large, as
is seen in the last time (see text for further details).\\~\\
Fig.4: log-log plots of the front velocity (lower curve) and width (upper
curve)
as a function of time in the inverse cascade process seen in Fig.3 in a system of size 
$2000$ and $\nu=1$.
 Both quantities
exhibit an initial exponential growth that turns to a power law growth (after
$t\approx 30$). The velocity is constant after this time, and the width
increases
like $t^\zeta$. Note that at the earliest time there is a slight decrease
in the
velocity; this is due to the decay of linearly stable modes that exist in random
initial conditions. \\~\\
Fig.5:
 The same as Fig.4 but with additive random noise for a system of size $1000$, 
$\nu=0.1$ and $ f=10^{-13}$. The velocity does not
saturate now, and the exponent $\zeta$ characterizing the increase of the width
with time changes to $\zeta=1.2 \pm 0.1$. The velocity increases in time like
$t^\gamma$ with $\gamma\approx 0.48 \pm 0.04$.\\~\\
Fig.6: log-log plot of the velocity as a function of system size for different
values of the noise amplitude,
${\sqrt f}=2.5\times10^{-13},2.5\times 10^{-9},2.5\times 10^{-7},10^{-6},
4\times 
10^{-6},2.5\times 10^{-5}$and $\nu =0.1$. For low values of $f$ we observe a
power law behavior, $v\sim L^\mu$, $\mu\approx 0.42 \pm 0.03$. For larger values
of $f$ there is a cross over to a stronger dependence on the system size,
see text
for details. \\~\\
Fig.7: The critical value of $L$ as a function of $f$ for which the
transition from
weaker to stronger dependence of $v$ on $L$ is observed, see Fig. 6.  In regime
I the noise is relatively small and the cusp picture is qualitatively
correct. In
regime II the noise is too large to leave the cusp picture intact, see
text  and  Figures  9 and 10  for
further details. The data indicates a power law as explained in the
text.\\~\\
Fig.8: log-log  plot of the velocity vs the noise amplitude for different
system sizes $\tilde L=5,10,40,160$, $\nu=0.1$.\\~\\
Fig.9: Typical flame fronts in regime I of Fig.7 where the system is sufficiently
small not to be terribly affected by the noise. The effect of noise in this
regime is
to add additional small cusps to the giant cusp. In figures a-d we present 
 fronts for growing system sizes  $\tilde L=10,20,40$ and $80$ respectively, 
$\nu = 0.1$ and ${\sqrt f}=2.5\times10^{-9}$.
 One can observe that when 
the system size grows there are more cusps with a more complex structure.\\~\\

Fig.10: A typical flame front in regime II of Fig.7. The system size is $160$,
 and the  noise amplitude is $2.5\times 10^{-5}$. This is sufficient to cause a
qualitative change in the appearance of the flame front: the noise introduces
significant levels of small scales structure in addition to the cusps. \\~\\

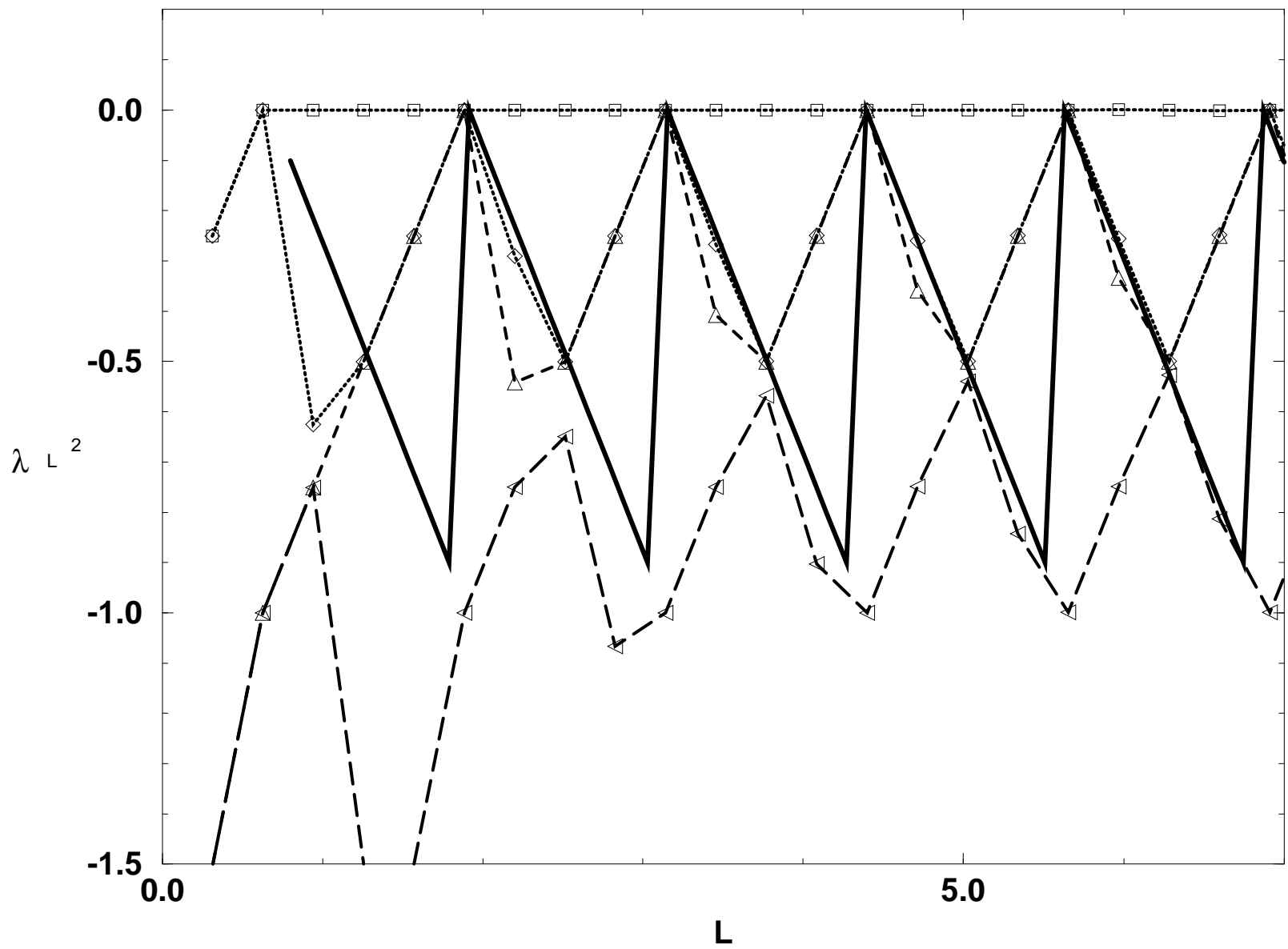
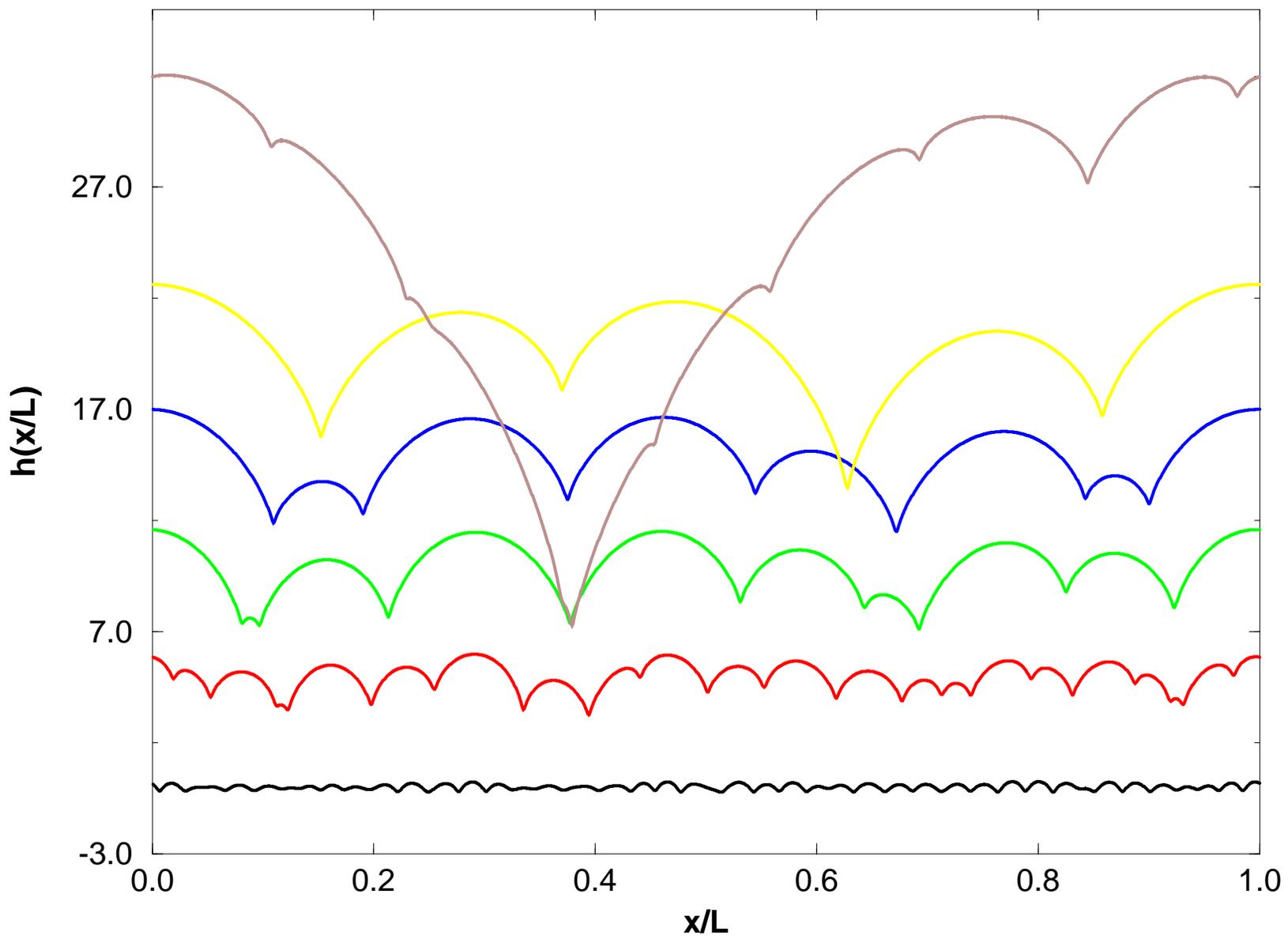
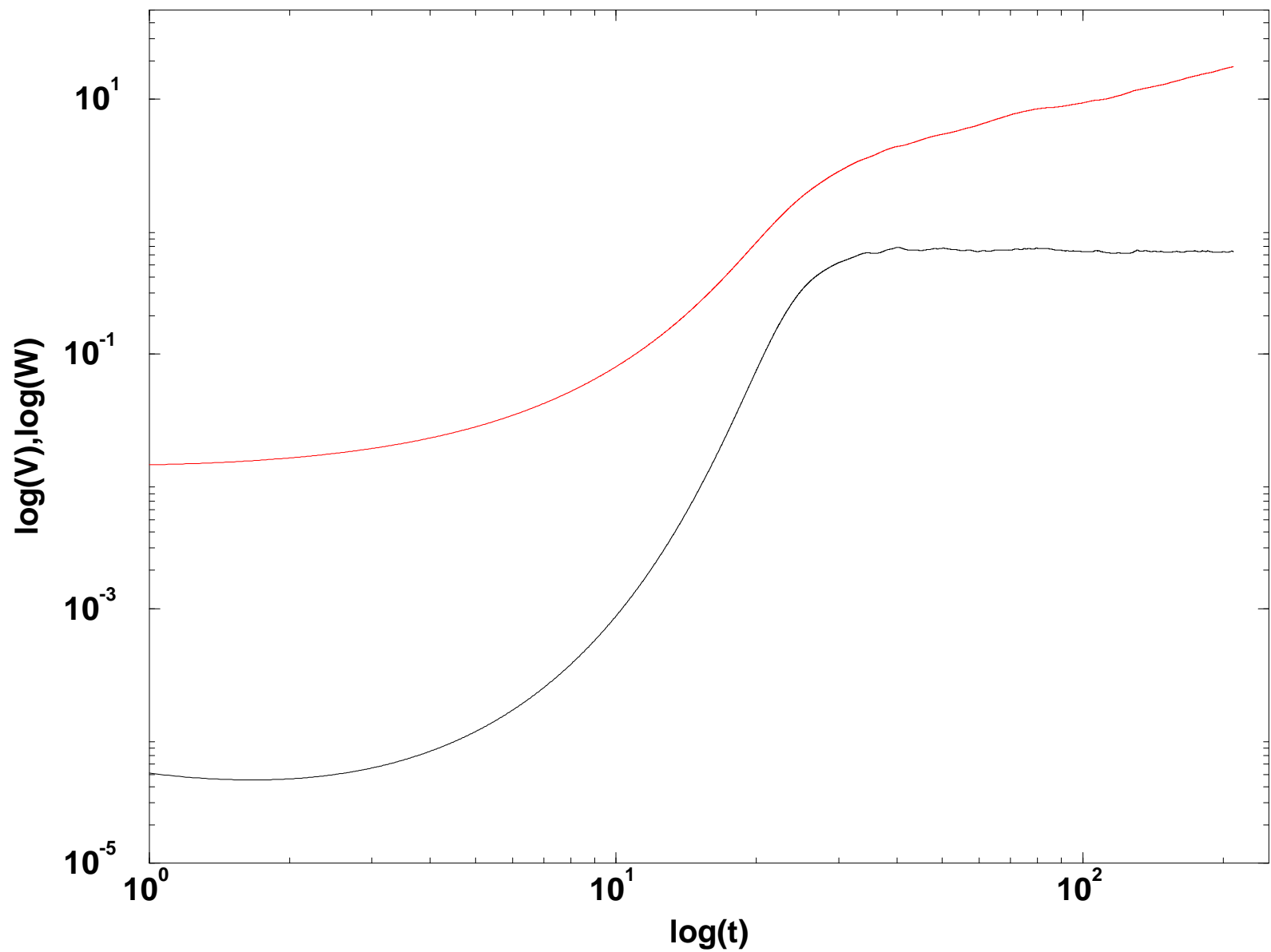
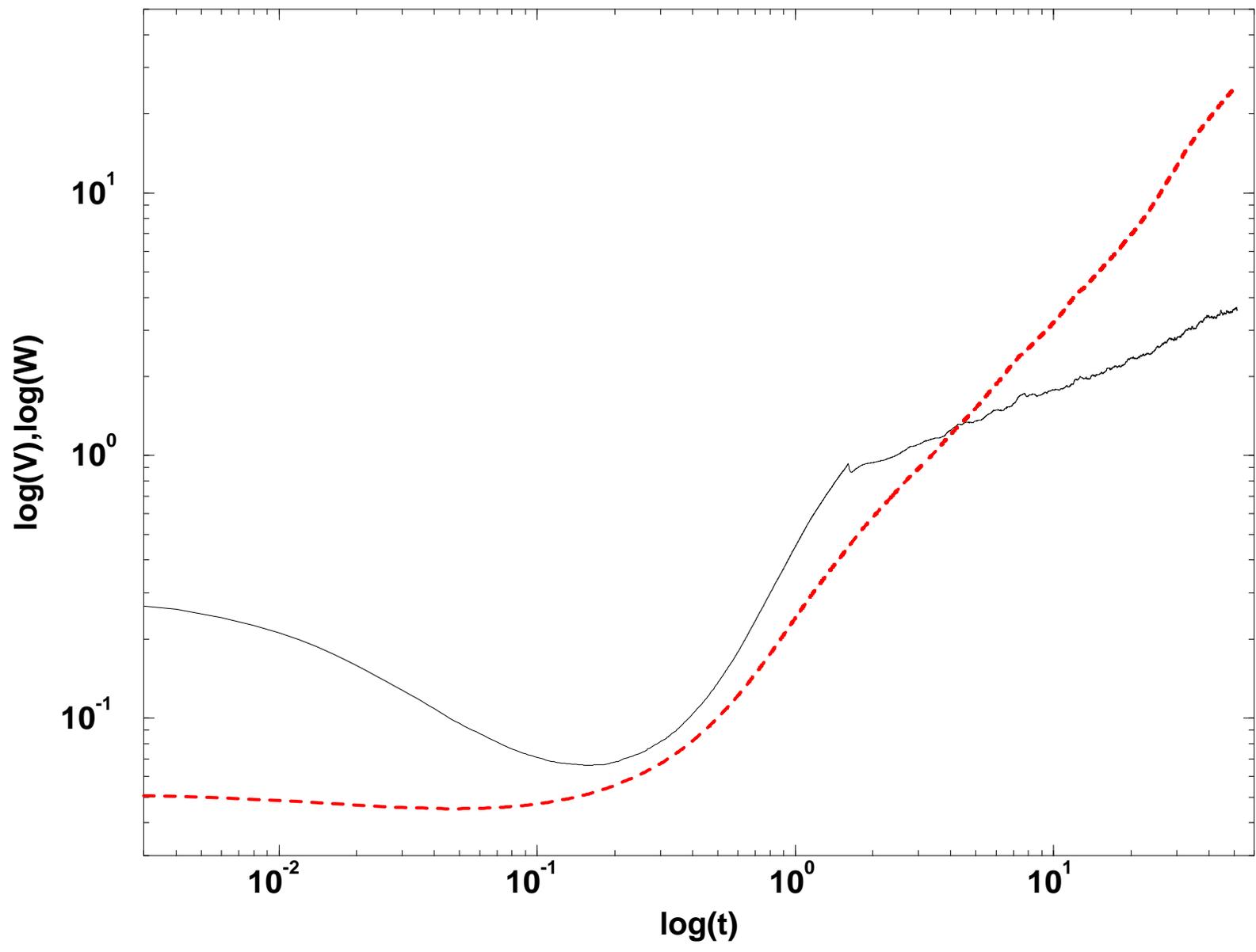
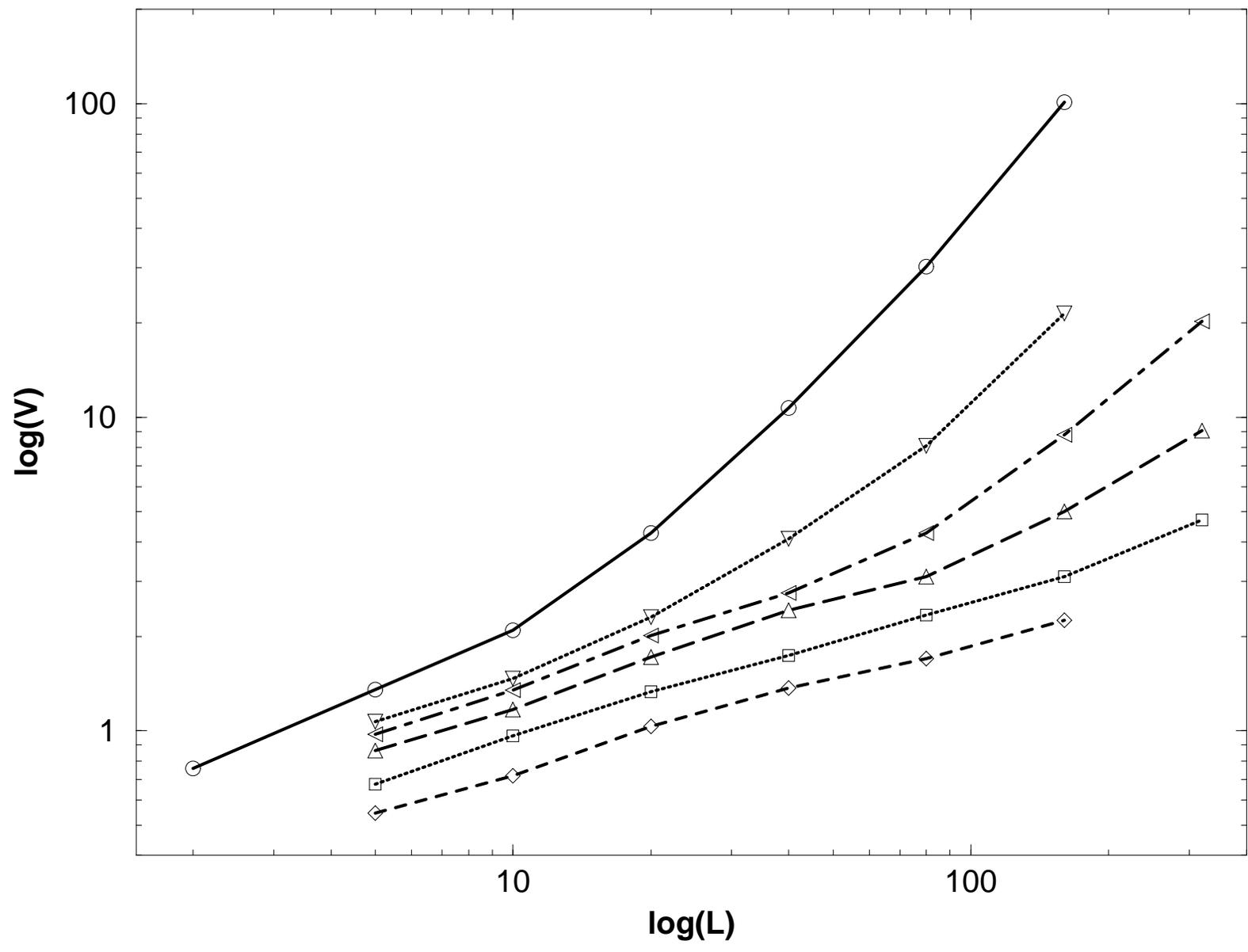
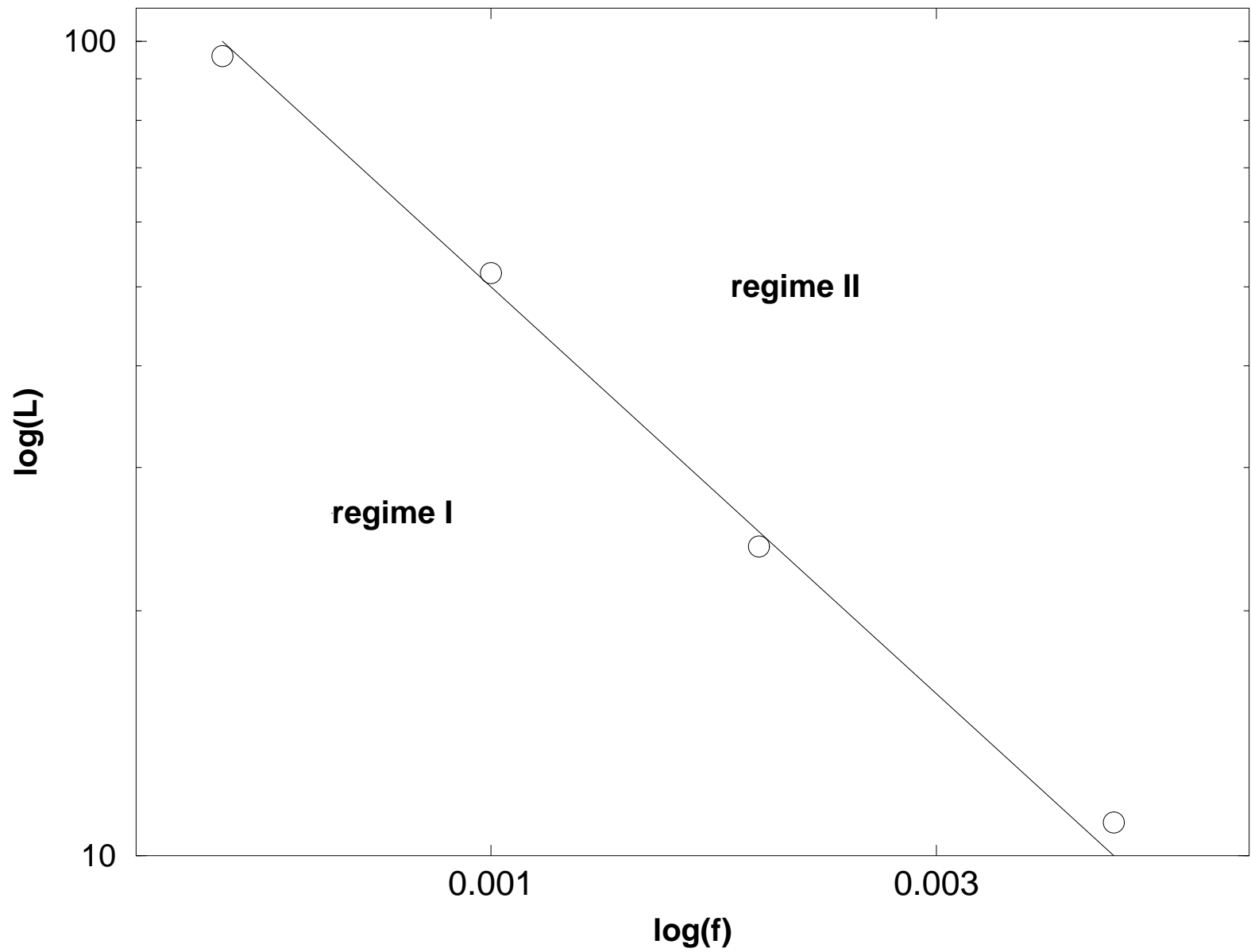
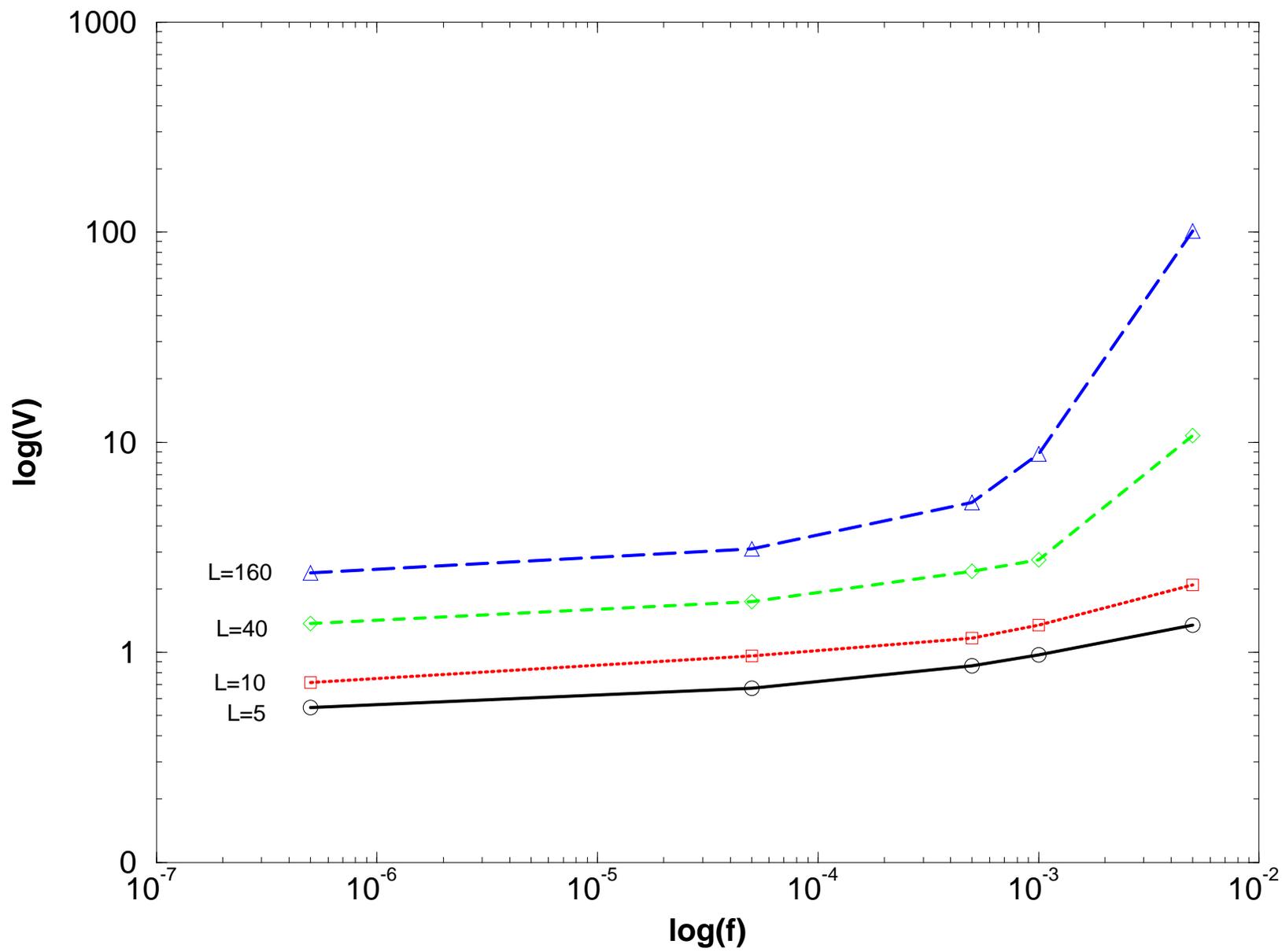
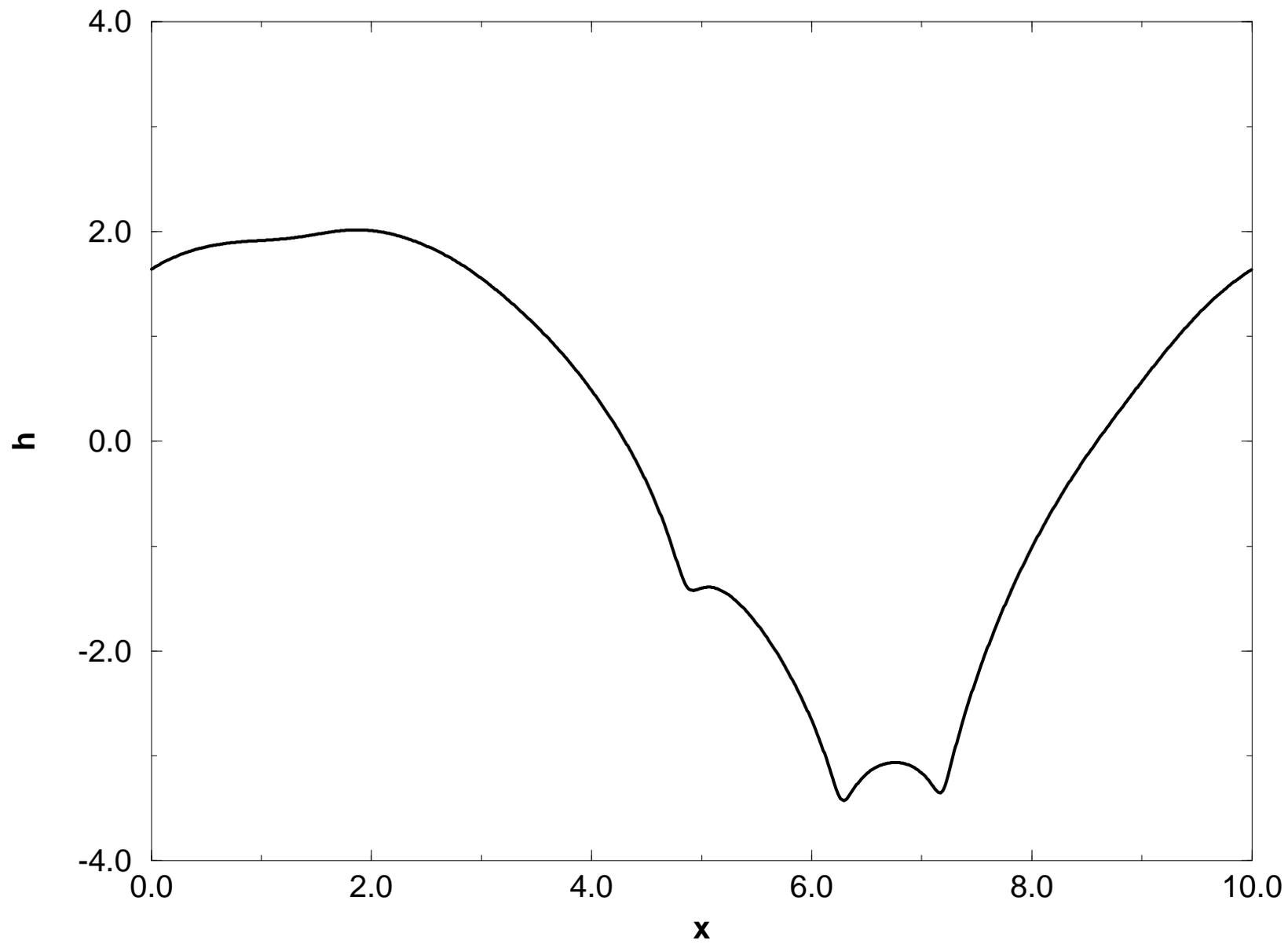
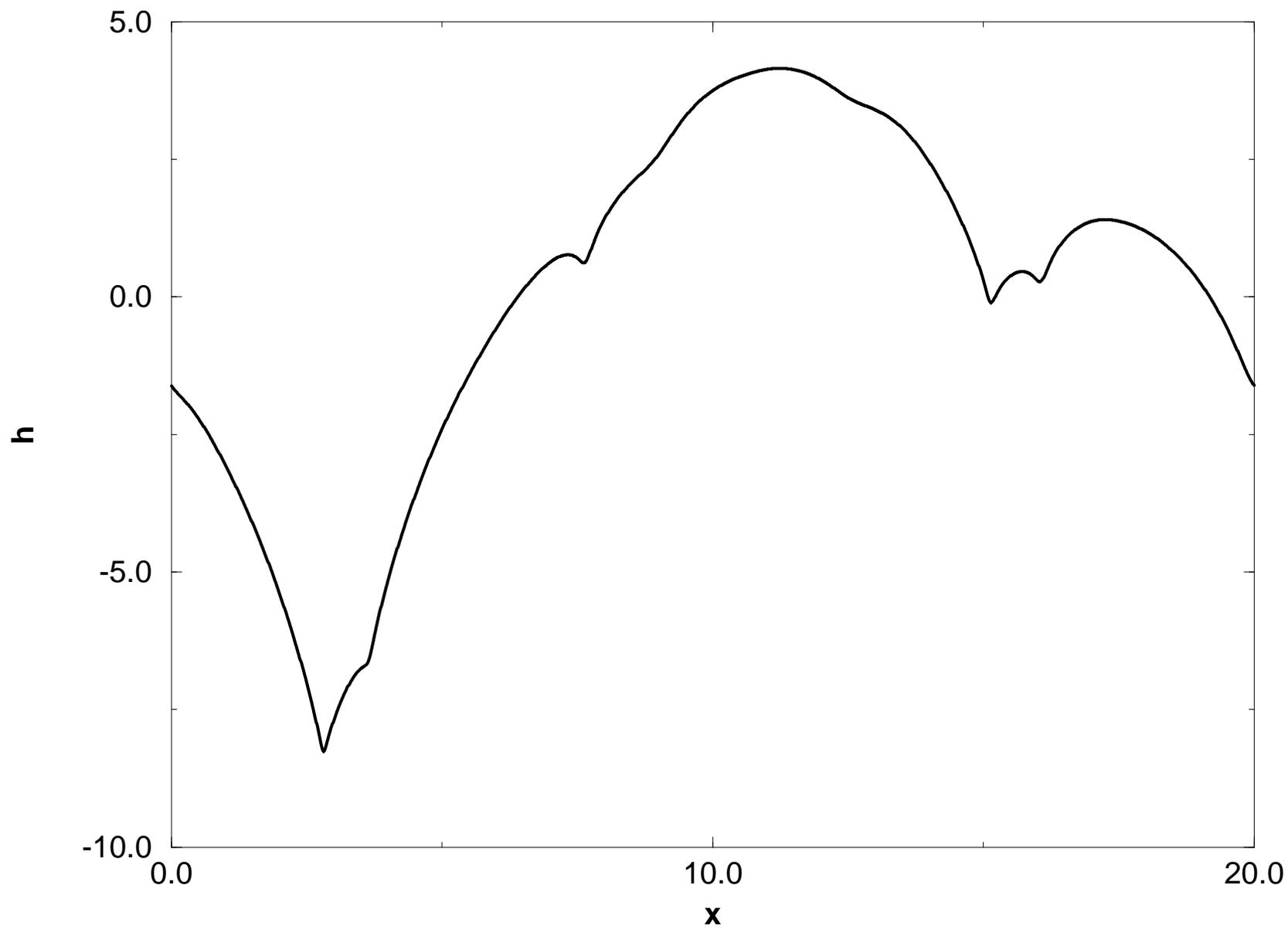
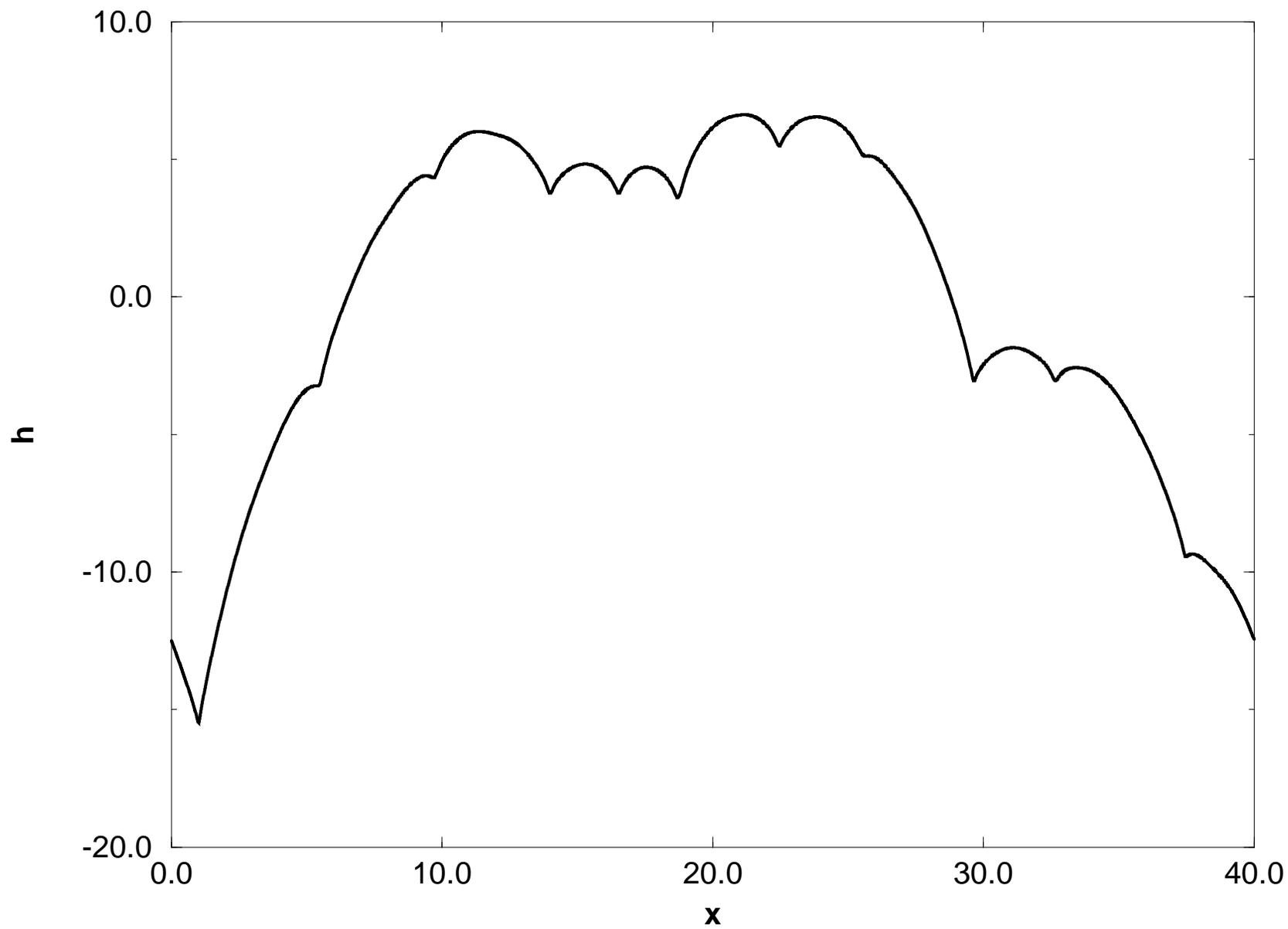
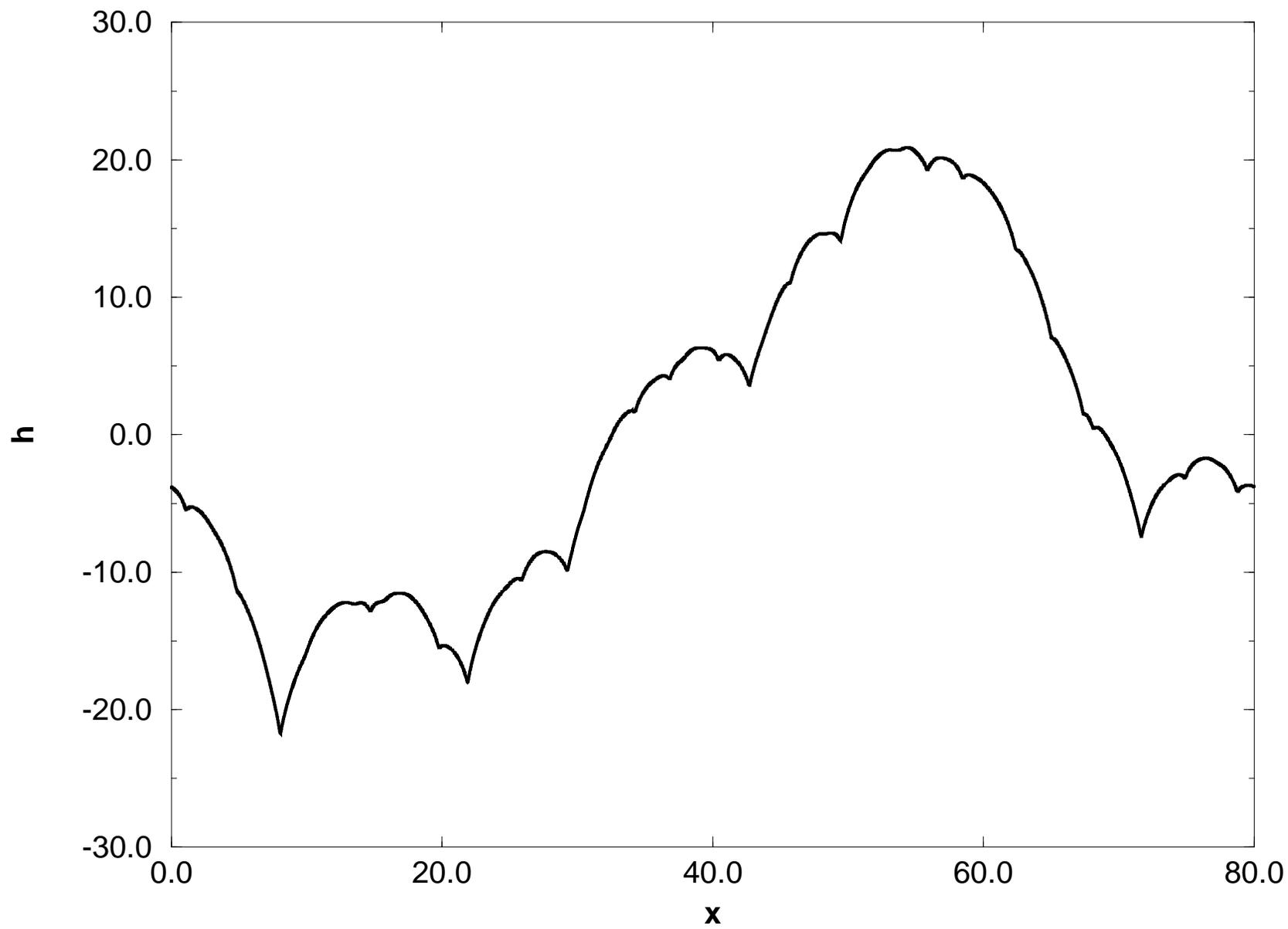
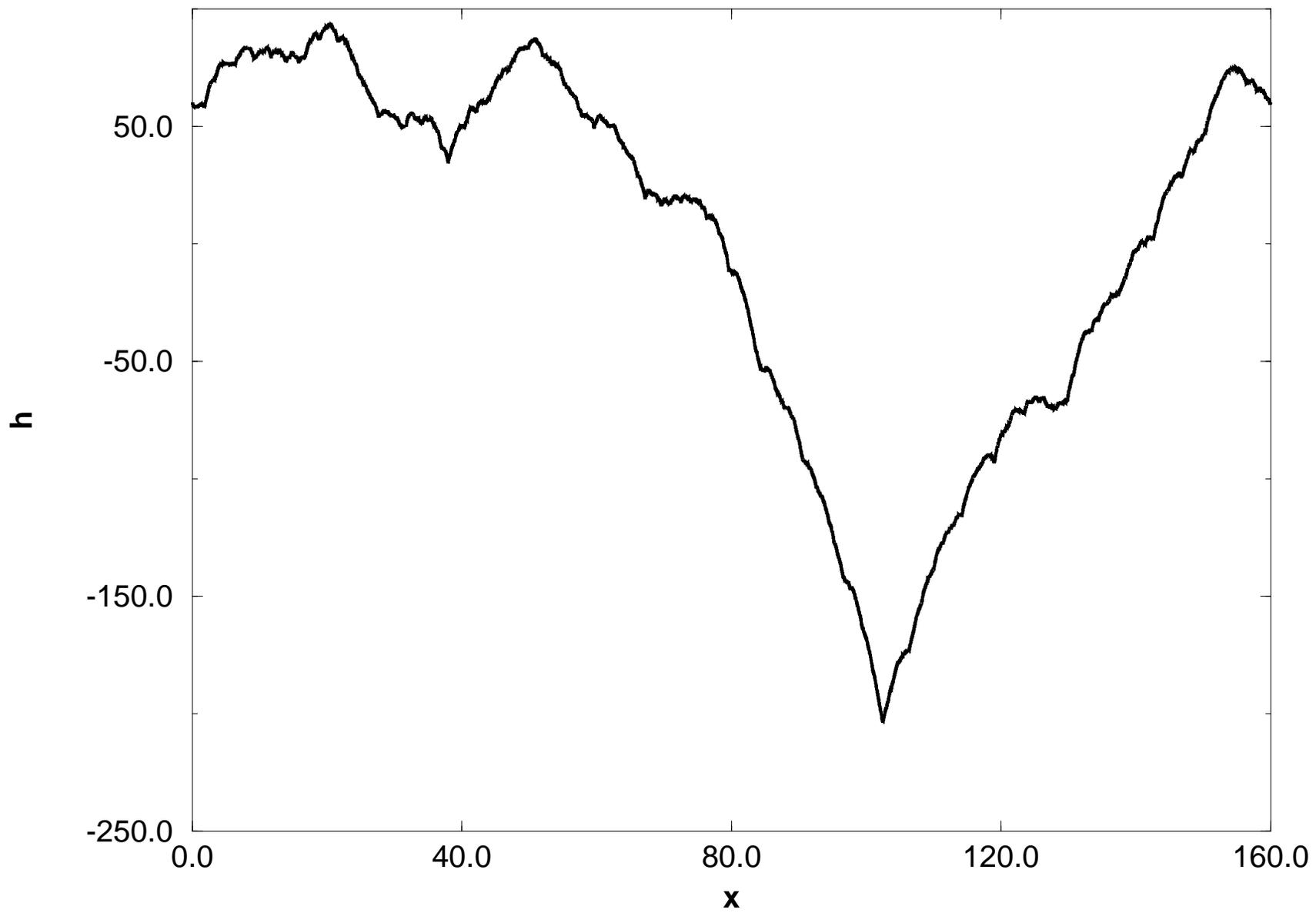
\end{document}